\documentclass[preprint,showpacs,preprintnumbers,amsmath,amssymb]{revtex4-1}
\usepackage{graphicx}
\usepackage{bm}
\usepackage{color}
\usepackage{amssymb}
\usepackage{epsfig}

\begin{document}

\title{Disorder perturbed Flat Bands I: Level density and Inverse Participation Ratio }
\author{Pragya Shukla}
\affiliation{Department of Physics,
 Indian Institute of Technology, Kharagpur, India.}
 
\date{\today}
 \widetext

 \begin{abstract}

We consider the effect of disorder on the tight-binding Hamiltonians with a flat band and derive a common mathematical formulation of the average density of states and inverse participation ratio applicable for a wide range of them. The system information in the formulation appears through a single parameter which plays an important role in search of the 
critical points for disorder driven transitions in flat bands \cite{psf2}. In weak disorder regime, the formulation indicates an insensitivity of the statistical measures to disorder strength, thus confirming the numerical results obtained by our as well as previous studies.

\end{abstract}

 

 \maketitle

\section{Introduction}

 Based on the dispersion relation,  the band structure of periodic lattices can in general contain two type of bands: often studied dispersive bands defined by energy $e$ as a function of Bloch wave-vector ${\bf k}$, and, the dispersion-less or flat bands defined by $e({\bf k})$=constant which appear under specific combinations of the system conditions. As indicated by  recent studies, the flat bands physics is not only of fundamental relevance, its detailed knowledge is significant from the industrial as well as technological view-point \cite{miel, tas, mits, ngk, ng1, hab, sr, hkv, vol, sh, cps, vid0, vidi, viddi, gu1, gu2}. The latter  has encouraged theoretical search of systems with flat bands, sometimes referred as  ''flatband engineering'' \cite{nmg1, nmg2, fl1, fl2, fl3, mapgf} as well as the analysis of system conditions e.g. role of symmetries in flat band existence and stability \cite{su, bg, gr, fl4}, the presence of magnetic field \cite{vid0}, the influence of disorder and interactions \cite{vidi, viddi, gu1, gu2}. Such bands are not mere theoretical models, they have  been observed in experimental studies too e.g. on photonic waveguides \cite{gmb, vcm, msc, mt, wmr, xhs}, exciton-polariton condensates \cite{myb, bgj, wcw} and ultarcold atomic condensates \cite{toi, jgt}.

The theoretical concept of a flat band is based on  the exact relations among a set of system conditions which may not  always be fulfilled in  a real solid. It is therefore relevant to seek the information about the effect of a weak perturbation  on the system condition e.g approximate symmetries, topological conditions, disorder on the flat band properties.
Due to highly degenerate nature of  the flat bands,
the response  to perturbations is expected to differ significantly based on  the location of the Fermi level i.e whether it is in the bulk of the flat or dispersive band, at the edge of a flat and dispersive band, or at the edge of two flat bands etc. Initial studies in this context, mostly numerical, have revealed a rich variety of behavior based on the nature of perturbation e.g. disorder and other system conditions (e.g. see \cite{nmg1, nmg2, sh, fl1, fl2, cps}) as well as the type of bands i.e single or many particle type \cite{vidi,viddi, gu1, gu2}. This motivates us to consider a theoretical  approach to study the response, based on the statistical analysis of a Hamiltonian with a generic combination of bands e.g a single or multiple flat bands, a flat band along with a  dispersive band etc. For a clear presentation of our ideas, here we confine the analysis to one specific perturbation, namely, disorder  with  primary focus on the single particle flat bands. Although the approach described here is in principle applicable to interacting flat bands too, it is technically complicated, requires a separate consideration and the detailed steps will be presented elsewhere. 

The presence of disorder leads to  randomization of the lattice-Hamiltonian and it can be best analyzed by an ensemble of its replicas. The choice of the appropriate ensemble is governed by the global constraints e.g. symmetries, conservation laws, dimensionality as well as local constraints e.g. disorder, hopping etc   and can be determined by the maximum entropy considerations.  
The underlying complexity however often conspires in favor of a  multi-parametric Gaussian ensemble as a good  model for many systems. For example, physical properties of complex systems in wide-ranging areas e.g. atoms, molecules, dynamical systems, human brain, financial markets  can be well-modeled by the stationary  Gaussian ensembles  if the underlying wave-dynamics is delocalized \cite{me, mj}  and by sparse Gaussian ensemble if the wave-dynamics is partially localized \cite{psand}. The success of these Gaussian models can usually be attributed to many independent sub-units contributing collectively to dynamics; the emergence of Gaussian behavior is then predicted by the central limit theorem. This encourages us to consider a flat band with Gaussian disorder with its Hamiltonian modeled by a multi-parametric Gaussian ensemble. As mentioned later in the text, the Gaussian consideration of disorder in case of a flat band has an additional technical justification too.

 Previous studies, based on theoretical as well as numerical analysis indicate that a multi-parametric evolution of the probability density of a Gaussian ensemble of Hermitian matrices, with arbitrary variances and mean values for its elements, can be expressed by a common mathematical formulation, governed by a single parameter \cite{ps-all}. The latter,  referred as the ensemble complexity parameter, is a function of all ensemble parameters and can act as a criteria  for the critical statistics \cite{psand}. In the present study,  we consider the complexity parameter formulation for a disorder perturbed flat band (also referred as disordered flat band) and derive the level density and inverse participation ratio in a generic form applicable for a wide 
range of such case.  Besides revealing interesting new features, these results are later used in  \cite{psf2} for the critical point analysis of the statistics of energy levels and eigenfunctions.

The paper is organized as follows. Our main objective here is to search for the criticality of the spectral statistics when a flat band is perturbed by the disorder. Due to technical complexity,   the theoretically analysis of this topic has not be carried out in  past (to best of our knowledge).   For a simple exposition of our ideas therefore,  here we primarily focus on the single particle bands perturbed by  Gaussian disorder. Section II.A briefly introduces a tight-binding periodic lattice with a flat band  along with a few well-known examples.  Onset of disorder  removes  the degeneracy of the flat band energy levels and affects their statistical correlations. An assumption of the Gaussian disorder, discussed in section II.B, permits us to model  the Hamiltonian by  a multi-parametric Gaussian ensemble in which the ensemble parameters i.e mean values and variances of the matrix elements depend on the system parameters e.g disorder, hopping, dimensionality etc. A variation of these parameters may subject the matrix elements to undergo a statistical evolution which can be shown to be governed by the ensemble complexity parameter \cite{psand, psco, psvalla, ps-all}.  This is briefly reviewed in section II.C along with the complexity parameters for the examples given in section II.A.  
As mentioned above, a flat band may arise under a wide range of system conditions including particle-particle interactions. Although technically complicated, the role of disorder in the flat bands caused by particle-interactions is an important topic which motivates us to include, in section III, a brief discussion of the complexity parameter formulation for these cases.  (A detailed investigation  of this topic requires a separate consideration and will be done elsewhere). Section IV reviews the complexity parameter formulation for the statistics of the eigenvalues and eigenfunctions for a multi-parametric Gaussian ensemble.  The information given in section IV  is used in section V to  derive the complexity parameter formulation of the density and the inverse participation ratio. As discussed in \cite{psf2} (part II of this work), a knowledge of these measures is necessary to seek the criticality in disordered flat bands. We conclude in section VI with a brief summary of our main results.

\section{Tight binding lattices with single particle flat bands}

\subsection{Clean limit}

Within tight-binding approximation, the Hamiltonian $H$ of a $d$-dimensional periodic lattice with ${\mathcal N}$ unit cells, each consisting of $M$ atoms,  with $\eta$ orbitals contributing for each atom, can be given as 
 \begin{eqnarray}
H = \sum_{x,y}   V_{xy} \; c_x^{\dagger} . c_y 
\label{h1}
\end{eqnarray}
with   
$c^{\dagger}_x, c_x$ as the particle creation and annihilation operators on the site  $x$ with 
$V_{xx}$ as the on-site energy and $V_{xy}$ as the  hopping between sites $x,y$. Here $x=({\bf n}, \alpha, \phi)$ where  ${\bf n}=(n_1,\ldots, n_d)$ are the indices for the $d$-dimensional unit cell, $\alpha$ is the atomic labels e.g. $\alpha=a,b$ for $M=2$  and $\phi=1, \ldots, \eta$ as the atomic orbital index. Hereafter the orbital index will be suppressed for the cases where only a single orbital from each atom contributes.

Due to periodicity of the lattice, the eigenstates $\psi$ of the Hamiltonian $H$ are delocalized Bloch waves with eigen-energies $e_{\nu}({\bf k})$ forming a band structure and ${\nu}$ as the band index: $\nu=1, \ldots, \mu$ with $\mu$ as the total number of bands.    
The nature of these bands  is sensitive to the system conditions (manifesting through $V_{xx}, V_{xy}$) which may give rise to dispersion-less bands defined by the energy $e_{\nu}({\bf k})=constant$  along with dispersive bands with their energy as a function of ${\bf k}$. The macroscopic degeneracy of the energy levels within flat band may lead to destructive interference of the Bloch waves, resulting in the localized or compact localized eigenstates (with zero amplitude outside a few unit cells) \cite{dr, drhmm, bg}.

\vspace{0.1in}

 Some prototypical examples can be described as follows:

\vspace{0.1in}

{\bf (a) 1-$d$ cross-stitch lattice with single orbital per site:} Referring the  unit cell by the label $m$, a site-index can be written as $x=(m, \alpha)$ with $\alpha=a,b$. The flat band in this case is obtained for following set of conditions:
(i) $ V_{xx}=0$, (ii) $V_{xy}=t$ for $x=(m,a), y= (m, b)$, (iii) $ V_{xy}=T$ if $x=(m, a)$ and $y= (m-1, \beta)$ or  $(m+1, \beta) $ with $\beta=a, b$, (iv) $V_{xy}=0$ for all other $x, y$ pairs; (see for example, \cite{fl3} for details)

\vspace{0.1in}

{\bf (b) triangular lattice with single orbital per site:} Again using the site-index $x=(m,\alpha)$ with $m$ as the unit cell label, the flat band condition can be described as  
(i) $ V_{xx}=2 t$ if $x=(m, a)$, (ii) $V_{xx}= \lambda^2 t$ if $x=(m, b)$, (iii) $V_{xy} =t$ if $x=(m, a), y=(m \pm 1, a)$, (iv) $ V_{xy}=T$ if $x=(m,a), y=(m, b)$ or $x=(m+1, a),  y=(m, b)$, 
(iv) $V_{xy}=0$ for all other $x, y$ pairs. (see Sec.5 of \cite{ngk} for the flat band conditions of this lattice).

\vspace{0.1in}

{\bf (c) 2-$d$-planer pyrochlore lattice with single orbital per site:} . 
 With 2-d unit cell labeled as $(m,n)$, one can write a site-index as $x=(m,n,\alpha)$ with $\alpha=a,b$ (i.e two atoms per unit cell). The lattice consists of one flat band   $ E_f = \varepsilon - 2 t$ and one dispersive band  $E_d=\varepsilon+2t (\cos k_x + \cos k_y +1)$ if $V_{xy}$ satisfies following set of conditions \cite{cps}:
(i) $V_{xx}=\epsilon$, (ii) $V_{xy}=t$ with $x=(m,n,\alpha)$ if $y=(m,n,\beta)$ or $(m-1,n,\beta)$ or $(m,n+1, \beta)$ with $\beta=a,b$ and (iii) $V_{xy}=0$ for all other $x,y$ pairs. 
(Note this case, with $\epsilon=2, t=1$, is used later for a numerical verification of our theoretical predictions).

\vspace{0.1in}

{\bf (d) 3-$d$ Diamond lattice with four fold degenerated orbitals on each site}

With 3-$d$ unit cell labeled as ${\bf r} \equiv (l,m,n)$, the site index can be written as  $x =({\bf r}, \alpha, \phi)$, $y =({\bf r'}, \beta, \phi')$ with  $\alpha, \beta=a,b$ and $\phi, \phi'=1,..., 4$. 
Here  hopping is considered between the orbitals within the nearest neighbor sites (on same or different unit cells ${\bf r}$ and ${\bf r'}$) with  
 (i) $V_{xy}=t_0$ if ${\bf r}={\bf r'}, \alpha=\beta$ and $\phi \not= \phi'$, 
 (ii) $V_{xy}=t_1$ if  $\alpha \not=\beta$ and $\phi = \phi'=1$, 
 (iii) $V_{xy}=t_2$ if $\alpha \not=\beta$ and $\phi =1, \phi'=4$, 
 (iv) $V_{xy}=t_3$ if $\alpha \not=\beta$ and $\phi =3, \phi'=4$, 
 (v) $V_{xy}=t_4$ if $\alpha \not=\beta$ and $\phi =2, \phi'=2$. As discussed in \cite{nmg2}, choosing 
$\varepsilon=0.0, t_0=0.0, t_1=-1, t_2=1, t_3=-1, t_4=-1$  leads to two flat bands \cite{nmg2}.

\vspace{0.1in}

\oddsidemargin=-70pt
\begin{figure}
\includegraphics[width=3cm,height=2cm]{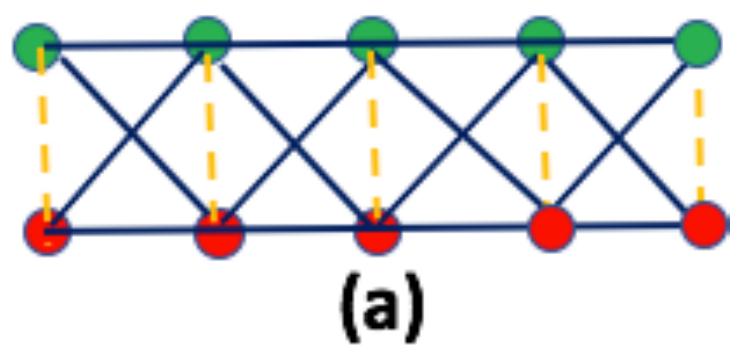}
\hspace{0.2in}
\includegraphics[width=3cm,height=2cm]{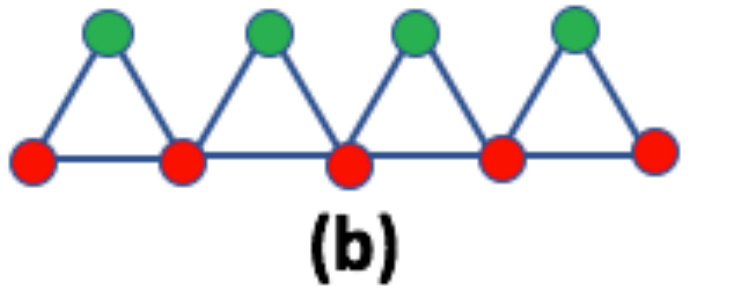}
\hspace{0.2in}
\includegraphics[width=3cm,height=2cm]{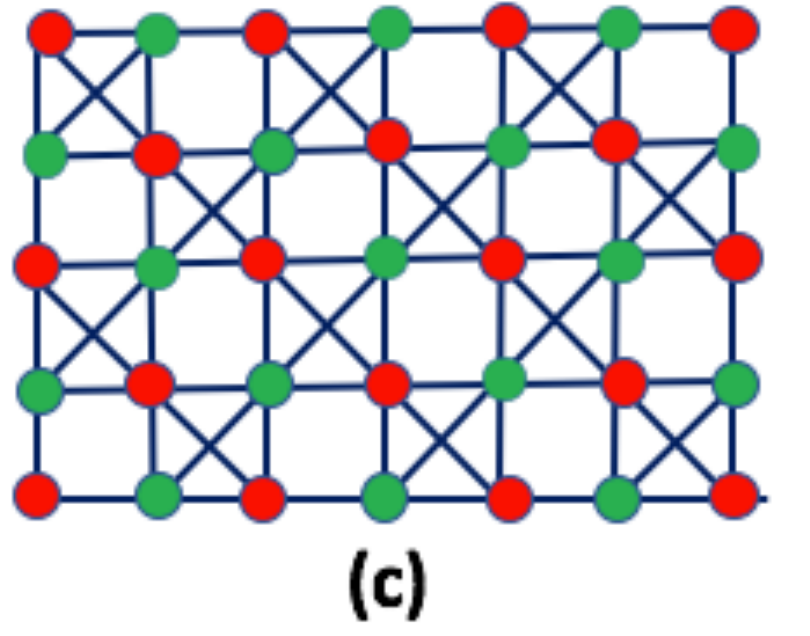}
\hspace{0.2in}
\includegraphics[width=3cm,height=2cm]{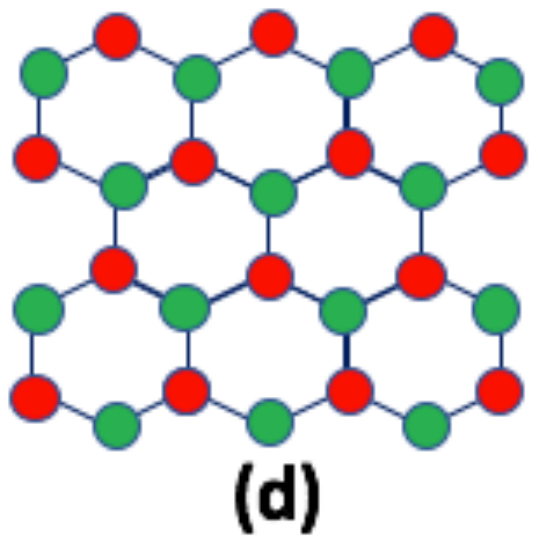}
\caption{
 (a) 1-$d$ cross-stitch lattice, 
(b) triangular lattice, 
(c) 2-$d$-planer pyrochlore lattice,
(d)  3-$d$ Diamond lattice.
}
\label{fig1}
\end{figure}

The examples mentioned above correspond to clean, single particle, bipartite lattices with time-reversal symmetry.  As an important application of the 
flat band studies is in context of magnetic systems, here we consider two 
example without time-reversal symmetry e.g. in presence of magnetic field: 

\vspace{0.1in}

{\bf (e) Aharonov-Bohm Cages} 

An important example giving rise to Aharonov-Bohm cages is the ${\mathcal T}_3$ lattice, a two-dimensional bipartite periodic structure with hexagonal symmetry and with three sites per unit cell (see figure 1 of \cite{viddi}). 
The presence of a magnetic field ${\bf B}$ affects the hopping element of $H$: $V_{xy}= t_{xy} \; {\rm e}^{i \eta_{xy}}$ with 
$\eta_{xy} = {2\pi \over \phi_0} \; \int_x^y {\bf A}.{\rm d} {\bf l}$ with ${\bf A}$ as the vector potential and $\phi_0=h c/e$ 
as the flux quantum.  Assuming a uniform magnetic field perpendicular to the plane of the lattice, the magnetic flux can be given as $\phi= B a^2 \sqrt{3}/2$ with $a$ as the lattice spacing.  For $\phi=0$ the spectrum has a flat band besides standard Bloch waves. But an unusual effect is caused by $\phi=\phi_0/2$, resulting in collapse of the energy spectrum  into three flat bands.  The high degeneracy of the energy levels in the  bands allows construction of the eigenstates localized in a finite size cluster, known as Aharonov Bohm cage;  (the term arises due to localization caused by Aharonov-Bohm type interference of electron-paths).  This case is discussed in \cite{viddi} in detail. 

Another simple system described by Hamiltonian in eq.(\ref{h1}) and leading to cage effect is a one dimensional chain of square loops with periodic boundary conditions kept in a uniform perpendicular magnetic field ${\bf B}$. This case is discussed in detail in \cite{vidi}.

\subsection{Effect of disorder}

As indicated by previous studies, the response of  a flat band is sensitive to the nature of disorder e.g. correlated vs uncorrelated and whether it causes a breaking of existing lattice symmetries \cite{fl2, fl4}. For example, a  randomization of the on-site energies leads to breaking of a chiral symmetry but the later is preserved if the hopping strengths are randomized \cite{fl4}. For clarity purposes,  the present study is confined to the randomized on site energies $V_{xx}$ only. The choice of an appropriate distribution for the latter depends on the available information and local system conditions. For cases with information only about first two moments of $V_{xx}$ (over an ensemble of disordered Hamiltonians), the maximum entropy hypothesis predicts a Gaussian distribution. The latter can also be justified on following grounds: 
due to macroscopic degeneracy of the levels, the density of states in the clean limit is a $\delta$-function which, in presence of a weak disorder, can be well-approximated by a limiting Gaussian distribution. An ensemble averaging of the density of states gives, by definition, the probability density of a typical energy state. Assuming the dominant contribution to energy states coming from the randomized on-site energies, the latter can then be appropriately described by a Gaussian. (Although the hopping strength also contributes to the energy states  but its effect is significant for the cases in which wave dynamics is extended in the unperturbed limit.  In the case of clean flat bands however most eigenfunctions are fully  or compact localized).   This motivates us to consider the case of a periodic lattice with  on-site uncorrelated Gaussian disorder;  the corresponding Hamiltonian is described by eq.(\ref{h1}) but with $V_{xx}$ as independent Gaussian  random variable.

To study the effect of on-site disorder, it is appropriate to represent $H$ in the site basis. For simplification, we now refer it as $|k \rangle$, $k=1 \to N$ with $N$ as the total number of sites. 
As the prototypical examples given in section II indicate, $H$ in the site basis  is in general a sparse Hermitian matrix, with degree of sparsity  governed by the dimensionality and range of hopping.  In presence of disorder however the effective sparsity may vary (based on relative strength of the non-zero elements) resulting in a change of behavior of the system with significant sample-dependent fluctuations. The joint probability distribution $\rho(H)$ of all independent matrix elements $H_{kl} \equiv \langle k| H | \rangle$, also referred as the ensemble density,  can then be given as  

\begin{eqnarray}
\rho(H) = C_w \; \prod_{k=1}^N {\rm e}^{-{(H_{kk}-V_{kk})^2 \over 2 w^2}}  \;\;  \prod_{k,l=cntd \atop q= 1\to \beta}^N \delta(H_{kl;q} - V_{kl;q})  \; \prod_{k,l \not=cntd \atop q=1 \to \beta}^N \delta(H_{kl;q} )  
\label{rho1}
\end{eqnarray}
with subscript $"q"$ of a variable referring to its real or imaginary component, 
$\beta$ as their total number ($\beta=1$ for real variable, $\beta=2$ for the complex one), 
 $C_w=\left({1\over \sqrt{2 \pi w^2}}\right)^{N} $ with the subscript $k,l=cntd$ refers to a pair of sites $k,l$ which are connected. Further representing the Dirac-delta function by its Gaussian limit i.e $\delta(H_{kl;q} - V_{kl;q})  \rightarrow    {\lim \atop {\sigma \to 0}} {1 \over \sqrt{2 \pi \sigma^2}} \; {\rm e}^{-{(H_{kl;q}-V_{kl;q})^2 \over 2 \sigma^2}}$, eq.(\ref{rho1}) can be rewritten as a multi-parametric Gaussian ensemble

\begin{eqnarray}
\rho(H)  = \lim_{\sigma \to 0}  \; C_{\sigma, w} \;  \prod_{k=1}^N {\rm e}^{-{(H_{kk}-V_{kk})^2 \over 2 w^2}}  \;\;  \prod_{k,l=cntd \atop q=1 \to \beta}^N {\rm e}^{-{(H_{kl;q}-V_{kl;q})^2 \over 2 \sigma^2}} \; \; \prod_{k,l \not=cntd \atop q=1 \to \beta}^N {\rm e}^{-{H_{kl;q}^2 \over 2 \sigma^2}} 
\label{rho2}
\end{eqnarray}

with $C_{\sigma, w}=\left({1\over \sqrt{2 \pi \sigma^2}}\right)^{N(N-1)} \; \left({1\over \sqrt{2 \pi w^2}}\right)^{N} $.

 On variation of the ensemble parameters, the ensemble density given by eq.(\ref{rho2}) is expected to undergo a multi-parametric evolution. But as shown in a series of studies \cite{ps-all, psvalla, psco},  the evolution  is indeed governed by a single parameter, a function of all ensemble parameters which is therefore referred as the ensemble complexity parameter. This is briefly reviewed in the next section.

\vspace{0.1in}

\subsection{Complexity parameter formulation}

  Consider an ensemble of Hermitian matrices $H$ with uncorrelated multi-parametric Gaussian density

\begin{eqnarray}
\rho (H,v,b)=C \; {\rm exp}[{-\sum_{q=1}^\beta \sum_{k\le l} {1 \over 2 v_{kl;q}} 
 (H_{kl;q}-b_{kl;q})^2 }]
\label{rho3}
\end{eqnarray}
with $C$ as the normalization constant, $v$  as the set of the variances 
$v_{kl;q}=\langle H^2_{kl;q}\rangle -\langle H_{kl;q} \rangle^2$ and $b$ as the set of all mean values $\langle H_{kl;q}\rangle=b_{kl;q}$.
Here the variances $v_{kl;q}$ and mean values $b_{kl;q}$ can take arbitrary values (e.g. $v_{kl;q} \to 0$ for non-random cases). It is easy to see that eq.(\ref{rho2}) is a special case of eq.(\ref{rho3}). Changing system parameters may lead to a variation of the ensemble parameters $v_{kl}, b_{kl}$ and a diffusion of the elements $H_{kl}$. But the evolution of $\rho(H)$ is described by a single parameter \cite{ps-all}


\begin{eqnarray}
{\partial \rho\over\partial Y} &=& \sum_{k,l; q}{\partial \over 
\partial H_{kl;q}}\left[{g_{kl}\over 2}
  {\partial \over \partial H_{kl;q}} +  \gamma \; H_{kl;q}\; \right] \rho 
\label{rhoy}
\end{eqnarray}

where $g_{kl}=1+ \delta_{kl}$ with $\delta_{kl}$ as a Kronecker delta function and 
 
\begin{eqnarray}
Y= -{1\over  \gamma \; N_{\beta}}  \; \; {\rm ln}\left[ \prod_{k \le l} \; \prod_{q=1}^{\beta}
 |1 - (2- \delta_{kl})  \gamma \; v_{kl;q}| \quad |b_{kl;q} + b_0|^2 \right] + constant.
\label{y1}
\end{eqnarray}
with $N_{\beta}={\beta N\over 2} (N+2-\beta) + N_b$ and  $N_b$ as the total number of $b_{kl;q}$ which are not zero. Further $b_0 =1$ or $0$ if $b_{kl;q}=0$ or $\not=0$, respectively and $\gamma$ as an arbitrary constant, marking the end of transition ($\rho(H) \propto {\rm e}^{-{\gamma \over 2} {\rm Tr} H^2}$ in steady state limit).

Eq.(\ref{rhoy}) describes a $Y$ governed Brownian dynamics of the elements $H_{kl}$ in the Hermitian matrix space, subjected to a Harmonic potential and is analogous to the Dyson's Brownian motion model \cite{me}. In the stationary limit $Y \to \infty$, the solution of the above equation corresponds to Gaussian orthogonal ensemble (GOE, for $\beta=1$) or Gaussian unitary ensemble (GUE, for $\beta=2$), the stationary universality classes of random matrices. A finite $Y$-solution of eq.(\ref{rhoy}) corresponds to a non-equilibrium state of crossover from an arbitrary initial ensemble to GOE or GUE and is referred as the Brownian ensemble (BE) \cite{me,apbe,pslg} intermediate between the specific initial condition $\to$ GOE /GUE.  It must be noted that eq.(\ref{rhoy}) has a unique solution for $Y \ge 0$.

A relevant initial condition, in context of  weakly disordered flat bands, is the Poisson spectral statistics, with localized eigenfunctions. The solution of eq.(\ref{rhoy}) for this initial condition is known as  Poisson $\to$ GOE/GUE Brownian ensemble; the latter is analogous to Rosenzweig-Porter (RP) ensemble (see \cite{rp, psbe, krav, psand} and references therein), a special case of eq.(\ref{rho3}) with 

\begin{eqnarray}
b_{kl;q} = \langle H_{kl;q} \rangle=0, \; \; \langle H^2_{kk} \rangle =1,\; \; \langle H^2_{kl;q} \rangle={1 \over  c \; N^{\gamma_0}}.
\label{rp1}
\end{eqnarray}
with $c$ and $\gamma_0$ as arbitrary  constants.
 From eq.(\ref{y1}), $Y$ for RP ensemble becomes
\begin{eqnarray}
Y = {1\over c \;  N^{\gamma_0}}.
\label{y0}
\end{eqnarray}
 For later reference, it is worth mentioning here that,  for $\gamma_0 =1, 2$, the statistical behavior of RP ensemble correspond to a critical level statistics with multifractal eigenstates \cite{psbe, psand, krav}.     A detailed study of RP ensembles is presented in \cite{psbe}. 
     
As eq.(\ref{rho2}) is a special case of eq.(\ref{rho3}),  $Y$ 
for a disordered lattice modeled by eq.(\ref{rho2}) can be obtained from eq.(\ref{y1}). 
For example, $Y(w, t, N)$, for the cases (a), (b), (c), (d) mentioned in section II, can be given as (for large $N$ and with $\beta=1$ for each case as $H$ is real-symmetric),

\begin{eqnarray}
Y &=&  -{1\over  \gamma \; N}  \; \; {\rm ln}\left[ 
 |1 -  \gamma \; w^2|^2 \quad |1-2 \gamma \; \sigma^2|^{(N-1)} \quad |t| \quad |T|^2 \right] 
+ const.\hspace{0.6in} {\rm  case (a)}   \label{y2}\\
&=& - {1\over  \gamma \; N}  \; \; {\rm ln}\left[ 
 |1 -  \gamma \; w^2|^2 \quad |1-2 \gamma \; \sigma^2|^{(N-1)} \quad (2 \; \lambda^4 \; t^6)  \right] + const
\hspace{0.5in} {\rm  case (b)}  \label{y3}\\
&=& -{1\over \gamma \; N}  \; \; {\rm ln}\left[ 
 |1 -  \gamma \; w^2|^2 \quad |1-2 \gamma \; \sigma^2|^{(N-1)} \quad |t|^3 \right] + const.  
\hspace{0.8in} {\rm  case (c)}  \label{y4} \\
&=& -{1\over  \gamma \; N}  \; \; {\rm ln}\left[ 
 |1 -  \gamma \; w^2|^2 \quad |1-2 \gamma \; \sigma^2|^{N-1}  (t_0 t_1 t_2 t_3)^2 \right] + const.  \hspace{0.6in} {\rm case (d)}
\label{y5} 
\end{eqnarray}

Similarly for the case(e), eq.(\ref{y1}) gives (with $\beta=2$)

\begin{eqnarray}
Y &=& - {1 \over \gamma \; N} {\rm ln} \left[ |1- \gamma \; w^2| \; |1-2 \gamma \sigma^2|^{(N-1)} \; |t|^{12}. |\cos\eta \sin \eta + \eta_0 |^3\right]  + const.  \hspace{0.14in} {\rm case (e)}
\label{y6}
\end{eqnarray}
with $\eta_0=1$  if $\sin\eta=0$ or $\cos\eta=0$, $\eta_0=0$ otherwise. 

Substitution of $w=0$ in eqs.(\ref{y2},\ref{y3},\ref{y4},\ref{y5},\ref{y6}) gives $Y$ for a clean flat band for each of the above cases. Here an important point worth notice is that $Y \propto {1\over N}$ for  cases (a)-(e) in large $N$-limit; as mentioned above, a similar dependence of $Y$ in RP ensemble for $\gamma_0=1$ correspond to a critical statistical behavior. But as discussed later in section IV,  this analogy by itself is not enough to predict the criticality of statistics for cases (a)-(e).   

 The applicability of the diffusion equation given above is not confined only to  a Gaussian potential (i.e Gaussian distributed matrix elements), it can be extended to the cases with generic single-well potentials too (see section 2 of \cite{psvalla}). Further, under generic conditions, the impurity distribution in the lattice may lead to pair-wise correlations among $H$-matrix elements e.g $\langle H_{ij;q} H_{kl;q} \rangle \not= 0$. Following maximum entropy hypothesis, $H$  can then be represented by an ensemble density
\begin{eqnarray}
\rho(H,a,b)  = C \; {\rm exp}\left[- \sum_{i,j,k,l; q}
 b_{ijkl;q}  H_{ij;q} H_{kl;q} - \sum_{k,l; q} a_{kl;q} H_{kl;q} \right],
\label{rhoc}
\end{eqnarray}
with $a_{kl;q}, b_{ijkl;q}$ as the distribution parameters and $C$ as a normalization constant. As discussed in \cite{psco}, the evolution of $\rho(H)$ in this case can again be described by eq.(\ref{rhoy}) but the form of $Y$ is now more complicated (see section II.A, eq.(1) and eq.(18) of \cite{psco}).

\section{Many-particle Flat bands} 

The tight binding approximation in eq.(\ref{h1}) is applicable for simple structures with single particle bands. As indicated by previous studies \cite{gu1, gu2}, flat bands can  also be generated or destroyed by turning on  the particle-particle interactions. For example, as discussed in \cite{gu2}, an interaction can be tuned in  a dispersive band structure of non-interacting electrons to yield an effective flat band of the interacting electrons. On the contrary, a flat band localization in Aharonov-Bohm cages, caused due to subtle interplay between structure geometry and magnetic field, is destroyed as soon as the particle interaction is switched on \cite{vidi}. It is then natural to seek the role of a disorder as a perturbation  of many particle flat bands. An important question in this context is  whether and how disorder and interactions compete with each other to cause a localization to delocalization transition and is this transition different from that in a dispersive band? 
Another relevant question is whether the presence of interactions can inhibit a critical spectral statistics or lead to multi-point criticality? In past, there have been attempts  to answer the above questions (see, for example, \cite{viddi, gu2}) but these are system-specific. It is therefore desirable to seek a complexity parametric formulation  which could be applicable to a wide range of many particle flat bands.
In this section, we briefly discuss the formulation for a simple interacting lattice.

\subsection{Lattice with $2$ electrons}

For a clear explanation, we first consider a simple case of $2$ electrons in a periodic lattice of ${\mathcal N}$ unit cells, with ${\mathcal S}$-sites per unit cell, described by the standard Hubbard Hamiltonian  
\begin{eqnarray}
H &=&     \sum_{k,l; s_1}  V_{kl} \; c_{k s_1}^\dagger  \; c_{l s_1}^{} +  \sum_{k, s_1, s_2}  U_k \; c_{k s_1}^\dagger \; c_{k s_1}^{} \; c_{k s_2}^\dagger \; c_{k s_2}^{}
\label{hi1}
\end{eqnarray}
with $|k \rangle$, $k=1 \to N$ as the $N$-dimensional site basis ($N ={\mathcal S} {\mathcal N}$) and $s_1, s_2 = \pm 1$ refers to two spin states of the electron. Here $U_k$ is the on-site interaction, $V_{kk}=\varepsilon_k$ is the on-site energy and $V_{kl}$ as the hopping element between sites $k, l$: $V_{kl} \not= 0$ if $k,l$ are neighboring sites and zero otherwise.

As examples, we consider the two dimensional ${\mathcal T}_3$ lattice or one dimensional chain of square loops (see case (e) of section II), placed in a uniform magnetic field of strength $B$ applied in direction perpendicular to the plane of the lattice.  
The latter results in a magnetic flux $f=e B a^2/2 h c$  through an elementary square ($a$ as the unit cell vector length, $h$ as the Planck's constant and $c$ as the speed of light) and the hopping element  $V_{kl}=t \; {\rm e}^{i \eta}$ with $\eta=2 \pi f$. (Note, for $U_k=0$, eq.(\ref{hi1})  corresponds to the Hamiltonian (\ref{h1})). As discussed in \cite{vidi} for zero on-site energies (i.e $V_{kk}=0$), the single particle spectrum consists of three bands with energies $\varepsilon_{\alpha}(k) = 2 \alpha \sqrt{1 + \cos(\eta/2) \cos(k a)}$ where $k \in \left[0, 2\pi/a \right]$, $\alpha= 0, \pm1$ is the band index. Clearly for $f \not=1/2$, only one single particle band is flat (with $\epsilon_0=0$) but for $f=1/2$ all three bands are flat. The latter in turn lead to five flat bands for two particle spectrum when $U_k=0$. Assuming the total spin polarization of the two electrons as zero (i.e with opposite spins) and a gauge in which only one hopping term per unit cell is modified for ${\bf B}\not=0$, the study in \cite{vidi} shows that, for $U_k=0$ and $\phi=\phi_0/2$, the lattice has non-dispersive flat bands with strong localization of  eigenstates but $U_k=U \not=0$ leads to their delocalization.

To understand the above behavior in terms of $Y$ based formulation, we proceed as follows.
For the matrix representation of $H$, we choose the anti-symmetrized two particle product basis   $|k s_1;  l s_2 \rangle = {1 \over \sqrt{2}} \left( |k s_1 \rangle . |l s_2 \rangle - |l s_2 \rangle . |k s_1 \rangle  \right) $ with $|k s_1 \rangle$ as the single particle state at the site $k$ for a particle with spin $s_1$. Clearly the product basis is  $N_2=N(N-1)/2$ dimensional and the matrix elements here satisfy following symmetries: 
$\langle i s_1; j s_2| H |k s_1; l s_2 \rangle  =  \langle j s_2; i s_1| H |l s_2; k s_1 \rangle$ 
and  $\langle i s_1; j s_2| H |k s_1; l s_2 \rangle = \langle k s_1; l s_2| H |i s_1; j s_2 \rangle$. 
From eq.(\ref{hi1}), with notation $ |\mu\rangle \equiv |k s_1;  l s_2 \rangle $ and 
$ |\nu \rangle \equiv |i s_3;  j s_4 \rangle$, a general matrix element $H_{\mu \nu}$ 
of $H$ in the product basis can then be written as
\begin{eqnarray}
H_{\mu \nu} &=&
\left(V_{ik} \; \delta_{jl} + V_{jl} \; \delta_{ik}\right)\delta_{s_1 s_3} \delta_{s_2 s_4} + 
 U_k  \; \delta_{ijkl} \; (1-\delta_{s_1 s_2})  
\left( \delta_{s_1 s_3} \delta_{s_2 s_4}-  
 \delta_{s_2 s_3} \delta_{s_1 s_4} \right)
 \label{h2mn}
\end{eqnarray}

with symbol $\delta_{ijkl}=1$ if $i=j=k=l$ and is zero otherwise. From eq.(\ref{h2mn}), $H_{\mu \mu} = V_{kk} + V_{ll} + U_k \; \delta_{kl} \; (1-\delta_{s_1 s_2})$ and $H_{\mu \nu}= t \; {\rm e}^{i \eta}$ if  $j=l$ and $i, k$  as nearest neighbor sites, or, 
$i=k$ with $j, l$ as nearest neighbor sites;  all other matrix elements are zero.  Clearly for non-zero on-site energies $V_{kk}$, the diagonals $H_{\mu \mu}$ need not be all uncorrelated and $H$, in general, can not be represented by an ensemble of type (\ref{rho3}). But for zero on-site energies and random, independent interaction parameters $U_k$, the diagonals are uncorrelated and can be random (if $k=l$, $s_1 \not=s_2$) or zero (if $s_1=s_2$). A choice of Gaussian distributed $U_k$ with mean $\langle U_k \rangle=U_0$ and variance $w^2$ then results in a same distribution for  $H_{\mu \mu}$ if $k=l, s_1 \not= s_2$. Further assuming non-random hopping i.e both $t$ and $\eta$ non-random, $H$ is a $N_2 \times N_2$ sparse complex Hermitian matrix with random diagonal elements and is  described  by the following ensemble density 

\begin{eqnarray}
\rho(H) = C_w \prod_{\mu \atop k=l, s_1 \not=s_2} {\rm e}^{-{(H_{\mu \mu}-U_0)^2 \over 2 w^2}}  \;\prod_{\mu; s_1=s_2} \delta(H_{\mu \mu})  \;  \prod_{\mu, \nu=cntd}^N \delta(H_{\mu \nu} -t \; {\rm e}^{i \eta})  \; \prod_{\mu, \nu \not=cntd}^N \delta(H_{\mu \nu} )  
\label{rho4}
\end{eqnarray}

Here the notation $\mu, \nu =cntd$ implies either sites $i, k$ and/or $j, l$ are connected by hopping $t$.  
Again describing the non-random  independent elements by a limiting Gaussian distribution, $H$ can be represented by the ensemble density in eq.(\ref{rho3}) but with indices $\{k,l\}$ now replaced by $\{\mu, \nu \}$ (thus $H_{kl; q} \rightarrow H_{\mu \nu; q}, v_{kl; q} \rightarrow v_{\mu \nu; q}, b_{kl; q} \rightarrow b_{\mu \nu; q}$).
\begin{eqnarray}
b_{\mu \nu} &=& t  \; {\rm e}^{i \eta} \hspace{0.5in}
{\rm for} \; (i,k= n.n \; and \; j=l) \; or \;  (j,l=n.n \; and \; i=k) \nonumber \\
b_{\mu \mu} &=&  U_0  \hspace{0.6in}  {\rm for} \; k=l, \; s_1 \not= s_2  
\label{b3}
\end{eqnarray}
Further, with only random $U_k$ and zero on-site energies,    
\begin{eqnarray}
v_{\mu\nu; q} = \langle (H_{\mu \nu} )^2\rangle-\langle H_{\mu \nu} \rangle^2  &=&
w^2 \; \delta_{q1} \; \hspace{0.5in} {\rm if \; i=j=k=l, \;  and  \; s_1 \not=s_2}, \nonumber   \\  
&=& \sigma^2  \hspace{1in} {\rm for \; other \; (\mu,\nu) \; pairs}.
\label{h3}
\end{eqnarray}

Substitution of eqs.({\ref{b3}, \ref{h3}) in eq.(\ref{y1}) gives, for $U_0 \not=0$, 

\begin{eqnarray}
Y \approx - {1 \over 2 \gamma \; N_2} {\rm ln} \left[  |1- \gamma \; w^2|^2 \; |1- \gamma \; \sigma^2|^2 \;|1-2 \gamma \; \sigma^2|^{(N_2-1)} \; |t|^{2 z} \; |\cos \eta \; \sin \eta +\eta_0|^z  \; U_0^2 \ \right]  + const. \nonumber \\
\label{yi1}
\end{eqnarray}

with $\eta_0$ same as in eq.(\ref{y6}) and $z$ as the number of nearest neighbors. As clear from the above, $Y$  now depends on the hopping parameters $t$ and $\eta$, disorder $w$ as well average strength $U_0$ of the interaction.  The above form of $Y$ can now be used to understand the localization tendencies of the eigenfunctions when interactions strengths $U_k$ become non-zero (e.g. the case discussed in \cite{viddi}). This however requires a theoretical formulation of the localization length or inverse participation ratio in terms of $Y$ which is discussed in section IV.B.

Note although the form of $Y$ in eq.(\ref{yi1}) is similar to eq.(\ref{y6}) but the statistical fluctuations at a given energy in the two cases can be significantly different. This is because a comparison of the fluctuations of two different systems require a prior rescaling leading to same background behavior on which fluctuations are imposed. As discussed later in section IV as well as in \cite{psf2}, the rescaling depends on the local mean level spacing which is in general different for non-interacting and interacting cases.

As mentioned above, the non-zero on-site energies (i.e $V_{kk} \not=0$) in eq.(\ref{hi1})  may in general lead to correlated diagonals. Assuming Gaussian distributed on-site energies, the ensemble density for such cases can again be described by eq.(\ref{rhoc}). As mentioned below eq.(\ref{rhoc}), $Y$ for these case can then be defined following the steps given in \cite{psco}.

\subsection{Lattice with more than two electrons}

The case discussed above corresponds to an electron density (number of electrons per site) $\eta_e=2/N$ which approaches zero in the thermodynamic limit $N \to \infty$. 
To analyze cases with finite electron density,  we now 
consider a $2$-body Hamiltonian $H$ representing dynamics of ${\mathcal M}$ interacting electrons in a periodic lattice of ${\mathcal N}$ unit cells, with ${\mathcal S}$-sites per unit cell. With $V$ and $U$ as single particle and two particle parts of $H$, one can write   

\begin{eqnarray}
H =\sum_{r,s=1}^N  V_{rs} \; c_{r}^\dagger \; c_{s}^{} +  {1\over 2} \; \sum_{r,s,t,u=1}^N U_{rstu} \; c_{r}^\dagger \; c_{s}^{} \; c_{u}^\dagger \; c_{t}^{}
\label{hint}
\end{eqnarray}
with $V_{r s}  \equiv \langle r | V | s \rangle$ and $U_{rsuv} \equiv \langle r s|U | u v \rangle$ and $|r \rangle, |s\rangle, |u \rangle, |v \rangle$ as the single particle states. Further assuming $p$ orbital per site, the total number of single particle states in the lattice are $N= p \; {\mathcal S} {\mathcal N}$. 
%
It is then appropriate to represent $H$ in a ${\mathcal M}$-particle Foch basis labeled by 
$\mu \equiv |n_1; n_2 \ldots n_N \rangle $, 
consisting of the occupation numbers $n_r$ of the single particle states labeled by $|r \rangle$ with $r =1 \to N$.  Due to two body selection rules, $H$ in the Foch  basis is a sparse matrix with total number of independent matrix elements as $N_2(N_2+1)/2$ where $N_2$ again corresponds to  the size of 2-particle basis space: $N_2=N(N-1)/2$. Thus $H$ in the ${\mathcal M}$-particle Foch basis is subjected to additional matrix constraints which in general lead to matrix element relations.

As an example, one can again consider the lattice with Hamiltonian in eq.(\ref{hi1}) but now with ${\mathcal M}$ electrons. With $\mu \equiv |n_{1s_1}; n_{2s_2} \ldots n_{Ns_N} \rangle $, $\nu \equiv |n'_{1s_1}; n'_{2s_2} \ldots n'_{Ns_N} \rangle $, the non-zero matrix elements for $H$ in this case are of following two types

(i) Diagonals $H_{\mu \mu}$
\begin{eqnarray}
H_{\mu \mu} &=& \sum_{r=1, s=\pm 1}^N \; n_{r s} \; \varepsilon_r + {1\over 2} \sum_{r=1}^N \; \sum_{s, s'=\pm 1}  n_{r s} \; n_{r s'} \; U_r \; (1-\delta_{ss'})
\label{a2} 
\end{eqnarray} 
 
(ii) off-diagonals $H_{\mu \nu}$ with  $n'_{ps}=n_{ps}+1, n'_{ts}=n_{ts}-1$, $n'_{rs} = n_{rs}$ $\forall r \not= p, t$
\begin{eqnarray}
H_{\nu \mu}  
&=& (-1)^{q'_p +q_t} \; (1-n_{ps})  \; n_{ts} \; V_{pt} \; \delta_{s s'}
\label{a3} 
\end{eqnarray} 
with $q_j = \sum_{k=1}^{j-1} n_{ks_k}$.   
Assuming  independent Gaussian distributed on-site energies with zero mean and variance $w^2$ and non-random nearest neighbor hopping $t$, eq.(\ref{a2}) and eq.(\ref{a3}) give the mean and variance of the non-zero matrix elements:
\begin{eqnarray}
& & b_{\mu \mu} = \langle H_{\mu \mu}\rangle = {1\over 2} \sum_{r=1}^N \; \sum_{s, s'=\pm 1}  n_{r s} \; n_{r s'} \; U_r \; (1-\delta_{ss'}), \nonumber \\
& & b_{\mu \nu} = \langle H_{\mu \nu}\rangle =  (-1)^{q'_p +q_t} \; (1-n_{ps})  \; n_{ts} \; V_{st} \; \delta_{ss'},\nonumber \\ 
& & v_{\mu \mu} = \langle H_{\mu \mu}^2\rangle - \langle H_{\mu \mu}\rangle^2 = w^2 \; \sum_{r=1,s=\pm 1}^N \; n_{rs}, \nonumber \\
& & v_{\mu \nu} = \langle H_{\mu \nu}^2\rangle - \langle H_{\mu \nu}\rangle^2 = 0.
\label{dis1}
\end{eqnarray}
Further, as clear from eq.(\ref{a2}), many diagonals may have contributions from a common set of variable $\varepsilon_r$ but are uncorrelated  i.e $\langle H_{\mu \mu} H_{\nu \nu} \rangle = \langle H_{\mu \mu} \rangle . \langle H_{\nu \nu} \rangle$ (for on-site energies $\varepsilon$ and interaction strengths $U_k$ as independent random variables with zero mean). 
The ensemble density $\rho(H)$ can then again be described by the real-symmetric version ($q=1$) of eq.(\ref{rhoc}) with indices $\{k,l \}$ replaced by $\{\mu, \nu \}$. The 
parameter $Y$  for this case can now  be obtained by substituting eqs.(\ref{dis1}) in eq.(\ref{y1}).

A technically useful point worth indicating here is the following. Although for clarity of presentation, we assumed a Gaussian distribution of on-site energies but it is not necessary. In fact, $H_{\mu \mu}$ being a sum over many independent random variables, the central limit theorem predicts it to be Gaussian distributed  for a wide range of the  distributions of on-site energies if $N$ is large. 

Although not relevant for our analysis, another point  worth mentioning here is that $H$-matrix in the many body Foch basis need  not be a two body random matrix ensemble (TBRME); the latter is defined as the one with all 2-body matrix elements  belonging to a Gaussian orthogonal ensemble (GOE) which requires the variances of all the diagonal same and two times that of the off-diagonals \cite{kota}). This is however not the case for  the Hamiltonian (\ref{hint}) which has many zero two body matrix elements due to finite range hopping.

\section{Diffusion of level density and inverse participation ratio}

The solution of eq.(\ref{rhoy}) for a desired initial condition at $Y=Y_0$ gives $Y$-dependence of the ensemble density.  By an appropriate integration, this can further be used to derive the $Y$-dependent formulation of the ensemble averaged measures; here we consider the formulation for the level density and the inverse participation ratio.

\subsection{Level density}


The $Y$-dependent ensemble averaged level density $R_1(e; Y)$ can be defined as $R_1(e; Y)=\sum_{n} \langle \delta(e-e_n) \rangle= \sum_{n=1}^N \int \delta(e-e_n) \; \rho(H; Y) \; {\rm d}H$. A direct integration of eq.(\ref{rhoy}) over $N-1$ eigenvalues and entire eigenvector space leads to an evolution equation for $R_1$ which occurs at a  scale $Y \sim N \Delta_e^2$ with $\Delta_e(e)$ as the local mean level spacing in a small energy-range around $e$: 

\begin{eqnarray}
{\partial R_1(e) \over\partial Y} = {\partial \over 
\partial e}\left[e - \int {\rm d}e' \; {R_1(e') \over e -e'} \right] R_1(e) 
\label{r10}
\end{eqnarray}

The solution of the above equation, also known as Dyson-Pastur equation, depends on the initial condition $R_1(e;0)$ and can be given as $R_1(e; Y)={1\over \pi} \; \lim_{\varepsilon \to 0} \; G(e-i  \varepsilon; Y)$ \cite{apps, apbe} where 

\begin{eqnarray}
G(z; Y) = G(z-Y G(z; Y); Y_0)
\label{g1}
\end{eqnarray}

For later reference, it must be noted that the limit $Y \to \infty$ corresponds to a semi-circle level density (expected as $\rho(H)$ approaches GOE/ GUE in the limit). But this limit is never reached if $R_1(e; Y_0)= \delta(e)$ \cite{apps}.

\subsection{Inverse Participation Ratio }

At the critical point, the fluctuations of eigenvalues are in general correlated with those of  the eigenfunctions. The  spectral features  at the criticality are therefore expected to manifest in the eigenfunction measures too.  As shown by previous studies \cite{mj}, this indeed occurs  through large fluctuations of their amplitudes at all length scales, and can be characterized by an infinite set of critical exponents related to the scaling of the ensemble averaged, generalized inverse participation ratio (IPR) i.e moments of the wave-function intensity  with system size.

The ensemble average of an  IPR  $\langle {\mathcal I}_q \rangle(e)$ at an energy $e$ is defined as $\langle {\mathcal I}_q \rangle(e) = \langle {\mathcal I}_q(e_k) \delta(e-e_k) \rangle$ with
 ${\mathcal I}_q(e_k) = \sum_n|\psi_{nk}|^{2q}$ as the IPR of the eigenfunction $\psi_k$ at the energy $e_k$, with components $\psi_{nk}$, $n =1 \to N$  in a $N$-dimensional discrete basis.
Clearly $ \langle {\mathcal I}_q \rangle(e)  \approx 1$ if the typical eigenfunctions in the neighborhood of energy $e$ are  localized on a single basis state, $\langle {\mathcal I}_q \rangle(e) =1/N^{q-1}$ for the wave dynamics  extended over all basis space. At transition, it reveals an anomalous scaling with size $N$: $\langle {\mathcal I}_q \rangle(e)  \sim N^{-(q-1) D_q/d}$ with $D_q$ as the generalized fractal dimension of the wave-function structure and $d$ as the physical dimension of the system. At the critical point,  $0 < D_q <d$, with $D_q$ as a non-trivial function of $q$.

For energy-ranges with almost constant level-density, it is useful to consider a local spectral average of IPR in units of the mean level spacing $\Delta(e)=(R_1(e))^{-1}$; it is defined as   
${\overline{\langle  I_q \rangle}} = {R_1(e)\over N} \; {\overline{\langle  {\mathcal I}_q \rangle}}$ where  
${\overline{\langle  {\mathcal I}_q \rangle}} = {1\over 2 D_e} \int_{e-D_e}^{e+D_e} \; {\langle {\mathcal I}_q \rangle} \; {\rm d}e $. As described in \cite{psbe, pslg}, the diffusion equation for  ${\overline{\langle  I_q \rangle}}$  can be given as 

\begin{eqnarray}
{\partial  {\overline{\langle  I_q \rangle}}  \over \partial \Lambda_I} 
\approx \left( a_q \; {\overline{\langle  I_{q-1} \rangle}} -   b_q \; {\overline{\langle I_q \rangle}} \right) + {E_c^2 \over 4 q N} \; \left[ \left(e + {2 N \over E_c} \right) {\partial \over \partial e} + {\partial^2 \over \partial e^2} \right]  {\overline{\langle  I_q \rangle}},  
\label{iq}
\end{eqnarray}
with 
\begin{eqnarray}
\Lambda_I = {4 N |Y-Y_0| \over E_c^2},
\label{almi}
\end{eqnarray}
and $a_q(e, Y)= {(2q-1) \overline{\langle u\rangle}  \over N}$, $b_q = 1+ {E_c^2 \over 4 q N}$ and $\overline{\langle u\rangle}(e; Y)$ as the ensemble averaged local intensity at energy $e$ and parameter $Y$. Here $E_c$ is an important system-specific energy-scale beyond which energy-levels are uncorrelated; it can usually be approximated by  the Thouless energy $E_{th}$. The latter corresponds to the energy scale that separates the GOE/GUE type of spectral behavior from system specific behavior \cite{krav}. For energy regime with fully localized and extended  dynamics, $E_{th} \sim \Delta_e$ and $O(N^0)$ but for partially localized regime 
it is believed to be $E_{th} \sim \Delta(e) \; N^{D_2/d}$ (with $\Delta(e)$ as the mean level spacing at energy $e$) \cite{krav}. For Rosenzweig-Porter model (eq.(\ref{rp1})), the study \cite{psbe} gives $E_{th} \propto N^{1-\gamma_0}$ (with $\Delta(e) \propto N^{-\gamma_0/2}$, $D_2 = (2-\gamma_0)/2$; note here $d=1$.

For energy regimes around $e$ where the approximation $e+{2N \over E_c} \approx {2N \over E_c}$ is valid, eq.(\ref{iq}) can be solved by the Fourier transform method. As in general $E_c \sim E_{th} \sim N^{-a}$ with $a \ge 0$,  the above approximation is usually valid for the bulk of the spectrum. 
To proceed further, we assume the validity of the above approximation and consider the Fourier transform $F_q(\omega) = \int   {\overline {\langle  I_q \rangle}}(e) \; {\rm e}^{i \omega e} \; {\rm d}e$ of eq.(\ref{iq}). This can be given as 

\begin{eqnarray}
F_q(\omega, \Lambda_I) \approx {\rm e}^{-\left(b_q-{i\omega E_c\over 2 q} +{\omega^2 E_c^2 \over 4 q N} \right) \; \Lambda_I} \left[ F_q(\omega,0)  +   \int_0^{\Lambda_I}   \; {\rm d}r \;   C_{q-1}(\omega, r)  \;  {\rm e}^{\;\left(b_q-{i\omega E_c\over 2 q} +{\omega^2 E_c^2 \over 4 q N} \right)\; r} \right] \nonumber \\
\label{iq0}
 \end{eqnarray}

where $C_q(\omega,r)=\int   {a_q \; {\langle  I_q \rangle}}(e, \Lambda_I) \; {\rm e}^{i \omega e} \; {\rm d}e$.
An inverse Fourier transform of eq.(\ref{iq0})    now leads to,   for $\Lambda_I >0$, 
\begin{eqnarray}
 {\overline {\langle  I_q  \rangle}}(e,\Lambda_I)  &=&  \sqrt{{ q N \over  \Lambda_I \; E_c^2}} \; \left[ {\rm e}^{-b_q \; \Lambda_I } \;  \int {\rm d}x \; {\rm e}^{-{q N \over \Lambda_I \; E_c^2} \; \left(e-x-{ \Lambda_I E_c\over 2 q}\right)^2} \;  {\overline {\langle  I_q  \rangle}}(x, 0) + \right . \nonumber \\
&+& \left . \int_0^{\Lambda_I} {\rm d}r   \int {\rm d}x \; {\rm e}^{-b_q \; (\Lambda_I-r)} \; 
{\rm e}^{-{q N \over  E_c^2 \Lambda_I}  \left(e-x-{(\Lambda_I-r)E_c\over 2q} \right)^2} 
\;  {\overline {\langle  I_{q-1}  \rangle}} (x, \Lambda_I) \; a_q(x)     \right]
\label{iq2}
 \end{eqnarray}

Based on the behavior of ${\overline {\langle  I_q  \rangle}}(x,0)$ and  ${\overline {\langle  I_{q-1}  \rangle}}(x,\Lambda_I)$,  the above equation can further be  reduced to a simple form. 
For example, for cases where both vary slower than the Gaussian in the integrals,  one can write

\begin{eqnarray}
 {\overline {\langle  I_q  \rangle}}(e,\Lambda_I)  &=&  \sqrt{ \pi} \; \left[ {\rm e}^{-b_q \; \Lambda_I } \;   {\tilde I}_q (0, 0) 
+ \int_0^{\Lambda_I} {\rm d}r  \; {\rm e}^{-b_q \; (\Lambda_I-r)} 
\; {\tilde I}_{q-1} (r, \Lambda_I) \; {\tilde a}_q(r)   \right]
\label{iq3i}
\end{eqnarray}
 where ${\tilde I}_q(r, \Lambda_I)  \equiv {\overline {\langle  I_q  \rangle}}  \left(e-{(\Lambda_I-r)E_c\over 2q}, \Lambda_I \right) $ and ${\tilde a}_q(r) =a_q\left(e-{(\Lambda_I-r)E_c\over 2q} \right)$. Further noting that main contribution to the integral in eq.(\ref{iq3i}) comes from the neighborhood of $r \sim \Lambda_I$, it can be approximated as 
 
\begin{eqnarray}
 {\overline {\langle  I_q  \rangle}}(e,\Lambda_I)  
&\approx &  \sqrt{ \pi} \; \left[ {\rm e}^{-b_q \; \Lambda_I } \;  {\tilde I}_q (0, 0) + 
\left(1- {\rm e}^{-b_q \; \Lambda_I} \right)
\;  {\tilde I}_{q-1} (\Lambda_I, \Lambda_I) \; {\tilde a}_q\left(\Lambda_I\right)   \right] 
\nonumber \\
&= &  \sqrt{ \pi} \; \left[ {\rm e}^{-b_q \; \Lambda_I } \; {\overline {\langle  I_q  \rangle}}(e,0) 
+ \left(1- {\rm e}^{-b_q \; \Lambda_I} \right) \; a_q(e) \;   {\overline {\langle  I_{q-1}  \rangle}}(e,\Lambda_I)  \right]
\label{iq3}
 \end{eqnarray}

The above approximation however is not applicable for the cases where $ {\overline {\langle  I_q  \rangle}}$ undergoes a rapid variation with energy. For example, with $
{\overline {\langle  I_q  \rangle}}(e,\Lambda_I){\overline = R_1(e, Y) \; {\langle  {\mathcal I}_q  \rangle}}(e,\Lambda_I)$, the initial condition ${\mathcal I}_q(e,0)= {\mathcal I}_0\; \delta_{e0}$ (for $q >1$) and $R_1(e,0)=N \delta(e)$ leads to 

\begin{eqnarray}
 {\overline {\langle  {\mathcal I}_q  \rangle}}(e,\Lambda_I)  &=&  
{1\over R_1} \; \sqrt{{ q N \over  \Lambda_I \; E_c^2}} \; \left[ N \; {\mathcal I}_0 \; {\rm e}^{-b_q \; \Lambda_I } \;  {\rm e}^{-{q N \over \Lambda_I \; E_c^2} \; \left(e-{ \Lambda_I E_c\over 2 q}\right)^2}  + \right . 
  \nonumber \\
&+& \left . \int_0^{\Lambda_I} {\rm d}r   \int {\rm d}x \; {\rm e}^{-b_q \; (\Lambda_I-r)} \; 
{\rm e}^{-{q N \over  E_c^2 \Lambda_I}  \left(e-x-{(\Lambda_I-r)E_c\over 2q} \right)^2} 
\;  {\overline {\langle  {\mathcal I}_{q-1}  \rangle}}(x, \Lambda_I) \; R_1(x) \; a_q(x)     \right] \nonumber \\
\label{iq4}
 \end{eqnarray}
with $R_1 \equiv R_1(e; \Lambda_I)$ outside the square bracket.

It is worth emphasizing here that eq.(\ref{iq2}) is applicable for an arbitrary dimension and band type (i.e dispersive or flat, single particle or many particle). But its may lead to different physical behavior based on $Y$ as  well as initial conditions. 


\section{Implications for weakly disordered flat bands}

In absence of disorder, the flat band is degenerate, say at energy $e=0$. The onset of  disorder, 
results in lifting of the degeneracy and an increase of the width of the flat band. 
For cases, in which disorder $w$ is the only parameter subjected to variation, eq.(\ref{y1}) gives 

\begin{eqnarray}
Y-Y_0  \approx {1\over \gamma \; N } \; {\ln|1-\gamma \; w^2|} , 
\label{yw}
\end{eqnarray} 
where $Y_0$ corresponds to the flat band with disorder $w=0$. With $Y$-parameters given by eqs.(\ref{y2}-\ref{y6}), the above result can easily be confirmed for cases (a)-(e) in section II. 
An important point worth indicating here is the following: $Y-Y_0$ is same irrespective of whether the disorder is varied in on site energy (as in cases (a)-(d)) or in the 2-body interaction (as in case (e)). Another point to note is that eq.(\ref{yw}) is not applicable if the interaction strength (if non-random) or its average value changes.   

Following from eq.(\ref{r10}) and eq.(\ref{iq}) along with eq.(\ref{yw}), a variation of disorder therefore leads to 
an evolution of  $R_1(e)$ as well as  $\langle I_2 \rangle$. As both these measure are needed to seek critical statistics, here we consider the solutions of  eq.(\ref{r10}) and eq.(\ref{iq}) in the weak disorder limit. For simplification and without loss of generality, we set $\gamma=1$.

\vspace{0.1in}

\subsection{Level Density}

A rescaling of $e \to {e \over w}$ in eq.(\ref{r10}), and the replacement $R_1(e) \to {N\over w} \; f_1\left(x\right)$ with $x={e\over w}$, leads to 
\begin{eqnarray}
{w|1-w^2| \over 2} \; {\partial f_1(x) \over\partial w} = {\partial \over 
\partial x}  \left[{w^2 \over N} \; x - \int {\rm d}y \; {f_1(y) \over x -y} \right] f_1(x). 
\label{px1}
\end{eqnarray}
For $w \ll 1$,  and in large $N$-limit, the above equation can further be approximated as 
$\left[ \int {\rm d}y \; {f_1(y) \over x -y} \right] \; f_1(x) \approx constant$ (neglecting  terms $ w \; {\partial f_1\over\partial w} $ and ${w^2 x\over N^2}$). The latter implies disorder as well as size independence of $f_1(x)$.

Alternatively, $R_1(e, Y)$ can directly be determined from eq.(\ref{g1}) as follows.
The level-density for a flat band in absence of disorder can be expressed as a $\delta$ function 
or its Gaussian limit 
$R_1(e; 0) =N \; \delta(e) =   \lim_{\sigma \to 0} {N\over\sqrt{2 \pi \sigma^2}} \; {\rm e}^{-{x^2 \over 2 \sigma^2}}$.
The initial condition on $G(z;Y)$ with $z=e-i \varepsilon$ then becomes   

\begin{eqnarray}
G(z; Y_0) =  \lim_{\sigma \to 0} \; {N \; \sqrt{2 \pi \sigma^2} \over \varepsilon \; e}  \; {\rm e}^{-{z^2 \over 2 \sigma^2}}
\label{g2}
\end{eqnarray}

For the above initial condition, eq.(\ref{g1}) gives  

\begin{eqnarray}
G(z; Y) =  \lim_{\sigma \to 0} \; {N \; \sqrt{2 \pi \sigma^2} \over \varepsilon \; e}  \; {\rm e}^{-{(z- (Y-Y_0) \; G)^2 \over 2 \sigma^2}}
\label{g3}
\end{eqnarray}
As  both $\sigma \to 0$ and $w \to 0$  in the above equation, $\sigma$ can be replaced by $w$ (note $\sigma$ is an arbitrary parameter in eq.(\ref{g2})). The solution of the above equation can then be given as (for $z$ satisfying $w^2 G^2, z G \ll {z^2 \over w^2}$)

\begin{eqnarray}
G(z; Y) = \lim_{w \to 0}   {N\; \sqrt{2 \pi w^2} \over \varepsilon \; e} \; {\rm e}^{-{z^2 \over 2 w^2 }} 
\label{g4}
\end{eqnarray}
The above in turn gives, for $w \ll 1$,

\begin{eqnarray}
R_1(e; w) =   {N\over \sqrt{2 \pi w^2}} \; {\rm e}^{-{e^2 \over 2 w^2 }}
\label{g5}
\end{eqnarray}

The above form of $R_1(e)$  is also confirmed by the numerical analysis of the two dimensional chequered board lattice with $N$ sites displayed in figure 2:  as shown in figure 2(b), a rescaling of $e$ by ${ w}$ results 
in  convergence of $R_1(e)$ behavior for different weak disorders to a single Gaussian curve.

For $w > 1$, although the left side of eq.(\ref{px1}) is no longer negligible, it is still satisfied, near $x \sim 0$,  by a solution of type $f(x) = {\rm e}^{-x^2/2}$. The latter implies the validity of eq.(\ref{g5}) for  $R_1(e)$ for $w >1$ too which is expected on the basis of analytical continuation of $R_1(e, w)$ from $w <1$ to $w >1$. A Gaussian behavior of the level-density for strong disorder is also predicted based on previous dispersive band studies of disordered systems.

\vspace{0.1in}

\subsection{Inverse participation ratio}

In absence of disorder, the flat band  can consist of localized states  and/or compact localized states \cite{dr, drhmm, bg}. The average IPR  for a flat band initial condition at $e=0$ and  $w=0$, equivalently $Y=Y_0$ or $\Lambda_I=0$ (see eq.(\ref{almi})), can then be written as  ${\overline{\langle {\mathcal I}_2 \rangle}}(e;Y_0) = {\mathcal I}_f \; \delta_{e0}$ with 
${\mathcal I}_f$ as the IPR of typical states at $e=0$   (with $\delta_{e0}=1$ or $0$ for $e=0$ and $e\not=0$, respectively).  
Further, with the normalization ${\overline {\langle  {\mathcal I}_1 \rangle}}(e, \Lambda_I) =1$ implying ${\overline {\langle  I_1 \rangle}}(e, \Lambda_I) ={R_1(e) \over N}$, eq.(\ref{iq3}) now gives, for $\Lambda_I \ge 0$,

\begin{eqnarray}
 {\overline {\langle  {\mathcal I}_2  \rangle}}(e,\Lambda_I)  &=& 
{1\over R_1} \; \sqrt{{ 2 N \over  \Lambda_I \; E_c^2}} \;\left[ N\; {\mathcal I}_f \; {\rm e}^{- \Lambda_I } \;  {\rm e}^{-{2 N \over \Lambda_I \; E_c^2} \; \left(e-{ \Lambda_I E_c\over 4}\right)^2}  +   J_0  \right]   \label{iq5} \\
J_0 &=&  {3 \over N} \; \int_{-\infty}^{\infty} J_1(e-x) \;  R_1(x) \; {\overline {\langle u \rangle}} (x)  
\label{j0}
 \end{eqnarray}

where $J_1(y) =\int_0^{\Lambda_I} {\rm d}r  \; {\rm e}^{- (\Lambda_I-r)} \; 
{\rm e}^{-{2 N \over  E_c^2 \Lambda_I}  \left(y-{(\Lambda_I-r)E_c\over 4} \right)^2} $. 
The latter can be expressed in terms of the Error function $\Phi$ (defined as $\Phi(u) = {2\over \sqrt{\pi}} \; \int_0^u  {\rm e}^{-x^2} \; {\rm d}x$),  

\begin{eqnarray}
J_1(y, \Lambda_I)  & = &    \sqrt{2 \pi \Lambda_I \over N} \; 
 {\rm e}^{-{4 y \over E_c} + {2 \Lambda_I \over N}} \; \left[
\Phi \left(\sqrt{N \Lambda_I \over 8} \left(1 +{4 \over N} -{ 4 y\over \Lambda_I E_c } \right)\right) - \Phi \left( \sqrt{2 \Lambda_I\over  N} \left(1 -{N y\over \Lambda_I E_c} \right) \right) \right] \nonumber \\
\label{j1}
 \end{eqnarray}

  For large $N$ and $\Lambda_I$, the above equation can further be approximated as
(using $\lim_{u \to 0} \Phi(u) =0$, $\lim_{u \to \infty} \Phi(u)= 1$ )

\begin{eqnarray}
J_1(y, \Lambda_I)   \approx    \sqrt{2 \pi \Lambda_I \over N} \; 
 {\rm e}^{-{4 y \over E_c} + {2 \Lambda_I \over N}} \; \Theta(y)
\label{j2}
 \end{eqnarray}

with $\Theta(y)$ as the step function: $\Theta(y) =0$  or $1$ for $y < 0$ and $y >0$, respectively.

As discussed in \cite{pslg},  the local intensity  ${\overline {\langle u \rangle}}$ at energy $e$ and parameter $Y$ for a Gaussian Brownian ensemble depends on its initial value at $Y=Y_0$. In case of a clean flat band at $e=0$ as an initial state at $Y=Y_0$,  the local intensity can be written as $\overline{\langle u\rangle}(e,Y_0) = u_0 \; \delta(e)$. Here $u_0$ is a constant, dependent on the state of localization of the eigenfunctions in the flat band. As discussed in \cite{pslg}, the $Y$ governed diffusion of the local intensity from this initial condition leads to $\overline{\langle u\rangle}(e,y) =  {u_0 \over \sqrt{2 \; \pi \; |Y-Y_0|}} \; {\rm exp} \left[-{e^2 \over {2 \; |Y-Y_0|}} \right]$ with $Y-Y_0$ given by eq.(\ref{yw}).
Further noting that $R_1(e)$ is a Gaussian too, both  $\overline{\langle u\rangle}$ as well as $R_1$ decay rapidly for $e \not=0$; as a consequence, the significant contribution to the integral over $x$ in eq.(\ref{iq5}) comes from the neighborhood of $x=0$. The integral $J_0$ can then be approximated as 

\begin{eqnarray}
J_0 \approx   \; {3\over N} \; J_1(e) \; \int_{-\infty}^{\infty}  {\rm d}x \;  R_1(x) \; {\overline {\langle u \rangle}} (x)  \approx { 3 \; u_0  \over w}  \; \sqrt{\pi \Lambda_I \over N} \; 
 {\rm e}^{-{4 e \over E_c} + {2 \Lambda_I \over N}} \; \Theta(e). 
\label{j3}
\end{eqnarray}

As a check, it is easy to see that, with $J_0$ given as above, the eq.(\ref{iq5}) gives the correct result for case $w=0$. 
Further analysis of eq.(\ref{iq5}) requires a prior knowledge of $\Lambda_I$ and $E_c$. 
Eq.(\ref{almi}) along with eq.(\ref{yw})  gives $\Lambda_I= {4 w^2 \over E_c^2}$ and eq.(\ref{g5}) gives $R_1(e) \approx {N \over \sqrt{2 \pi w^2}}$ near $e \sim 0$.
The first term of eq.(\ref{iq5}) then rapidly decays for $e \not= {w^2 \over E_c}$ and/or for $w \ge E_c$. 
Assuming $E_c \sim  N^{-\mu}$ with $\mu >0$, eq.(\ref{iq5})  can now be approximated as 
\begin{eqnarray}
{\overline {\langle  {\mathcal I}_2  \rangle}}(e \sim 0)  & \approx &  
{6 \; \pi\; u_0 \over N \; E_c} \;  
 {\rm e}^{{8 w^2 \over N E_c^2}}  
\label{aqq}
\end{eqnarray}
In general, $E_c$  is disorder dependent and  $\mu$ can vary with $w$. Thus $ {\overline {\langle  I_2  \rangle}}$  for large $w$ can in general depend on both disorder as well as size. For small-$w$, however,  the approximation ${\rm e}^{8 w^2 \over N E_c^2} \sim 1$  (valid for  $w <  N^{(1-2\mu)/2}$, assuming disorder independence of $E_c$ for weak disorder) leads to  a disorder-independent average IPR at $e \sim 0$
\begin{eqnarray}
{\overline {\langle  {\mathcal I}_2  \rangle}}(e \sim 0)  & \approx &  
{6 \; \pi\; u_0 \over N \; E_c} \;   \hspace{0.5in}    {\rm for} \; w< N^{(1-2\mu)/2}
\label{aq9}
\end{eqnarray}

For $e > 0$, the energy-dependence of eq.(\ref{iq5}) as well $R_1(e)$ can no longer be neglected. 
It now leads to 
\begin{eqnarray}
{\overline {\langle  {\mathcal I}_2  \rangle}}(e )  & \approx &  
{6 \; \pi \; u_0 \over N \; E_c} \;  
 {\rm e}^{{8 w^2 \over N E_c^2}} \;  {\rm e}^{-{4 e \over E_c} +{e^2 \over 2 w^2}}
\label{aq10}
 \end{eqnarray}
As clear from the above, average IPR now decays exponentially with increasing energy (with $E_c \sim N^{-\mu}$ with $\mu  >0$).

At this stage, it it relevant to know the size-dependence of $E_c$. 
As mentioned in section IV B, $E_c \sim E_{th} \sim \Delta(e). N^{D_2/d}$ in partially localized regime. The numerics for two dimensional Chequered board lattice  (case (c) in section II, with a flat band at $e=0$ for $\epsilon=2, t=1$) suggests $D_2 \approx 1$ (see figure 4); with $d=2$ and  $\Delta(e) =R_1^{-1}(e) \propto N^{-1}$ which  gives $E_{th} \propto N^{-1/2}$. Substitution of the latter  in eq.(\ref{aq9}) then  gives  ${\overline {\langle  {\mathcal I}_2 \rangle}} \propto N^{-1/2}$ which is consistent with our numerical analysis  (see figure 3.c). 

For $w >1$, eq.(\ref{yw}) gives $Y-Y_0 =  -{ \ln |1-w^2| \over N}$. As mentioned above, the form of $R_1(e)$ for $w>1$ in case of a single perturbed flat band is same as that of $w <1$. 
With $\Lambda_I =  {4 \ln |1-w^2| \over E_c^2}$, eq.(\ref{iq5}) now gives
\begin{eqnarray}
{\overline {\langle  {\mathcal I}_2  \rangle}}(e )  & \approx &  
{6 \; \pi \; u_0 \over N \; E_c} \;  
 {\rm e}^{{8 \ln |1-w^2| \over N E_c^2}} \;  {\rm e}^{-{4 e \over E_c} +{e^2 \over 2 w^2}}
\label{aq10}
 \end{eqnarray}
But $E_c$ being disorder dependent,  $\mu$ for $w >1$ need not be same as that of $w <1$. Thus $ {\overline {\langle  I_2  \rangle}}$  for $ w>1$ can in general depend on disorder and energy.

\section{ Role of other bands in the vicinity}

In general, a clean system may contain more than one flat band as well as dispersive bands. Although for weak disorder, the neighborhood has negligible influence on bulk of the flat band, the strong disorder leads to its spreading and overlap with other bands. 
For example, for a dispersive band in the vicinity, it may lead to an increase of the dispersive level density at the cost of flat band one, eventually  leading to a merging of the bands. If the neighborhood consists of another flat band separated by a gap, increasing disorder would lead to a rise of the level density in the gap-region followed by a merging of the Gaussian densities.  As discussed below, this may also affects the average behavior of level density and IPR as well as  their fluctuations.  

\subsection{Level Density}

As $R_1(e;Y)$ given by eq.(\ref{g5}) is derived for a $\delta$-function initial condition, it is applicable only for an isolated flat band. In presence of other bands in the neighborhood, eq.(\ref{r10}) should be solved for the altered initial conditions.
As clear from the integral in eq.(\ref{r10}), $R_1(e;Y)$ at energy $e$ can be affected by other parts of the spectrum, say $e' \not=e$ if $R_1(e')$ is very large. For weak disorder cases, therefore it is appropriate to solve eq.(\ref{r10}) with an initial condition  $R_1(e,0)$ valid for all $e$ ranges. 
Here we consider two examples: 

(i)  {\it two flat bands}: For this case, we have  $R_1(e;Y_0)= {N\over 2} \sum_{k=1}^2 \;  \delta(e-e_k)$ with $e_1, e_2$ as the band-locations, satisfying the normalization condition $\int_{-\infty}^{\infty} R_1(e) \; {\rm d}e =N$. The initial condition on $G$ can now be written as 
$G(z; Y_0) =  \lim_{\sigma \to 0} \; {N \; \sqrt{2 \pi \sigma^2} \over  2\; \varepsilon \; e} \sum_{k=1}^2 \; {\rm e}^{-{z_k^2 \over 2 \sigma^2}}$ where $z_k= z-e_k$ with $z=e - i \varepsilon$. Eq.(\ref{g1}) then gives  

\begin{eqnarray}
G(z; Y) =  \lim_{\sigma \to 0} \; {N \; \sqrt{2 \pi \sigma^2} \over  2 \; \varepsilon \; e}  \; \sum_{k=1}^2 {\rm e}^{-{(z_k- |Y-Y_0| \; G)^2 \over 2 \sigma^2}}
\label{g3}
\end{eqnarray}

Using the approximations $w^2 G^2, z G \ll {z^2 \over w^2} $ and proceeding as in the single band case, it can again be shown that in weak disorder limit  

\begin{eqnarray}
R_1(e; w) =   {N\over 2 \; \sqrt{2 \pi w^2}} \; \sum_{k=1}^2 \; {\rm e}^{-{(e-e_1)^2 \over 2 w^2} }  
\label{gf3r}
\end{eqnarray}
For later reference, it is instructive to look at $R_1$ behavior near $e \sim (e_1+e_2)/2$:
\begin{eqnarray}
R_1\left({e_1+e_2\over 2}; w \right) =   { N\over \sqrt{2 \pi w^2}} \; {\rm e}^{-{(e_2-e_1)^2 \over 8 w^2} }  
\label{gf3r1}
\end{eqnarray}

Clearly the gap $|e_1-e_2|$ is increasingly filled up  with levels  as disorder increases (due to level repulsion) and the Gaussians start merging for $w > |e_1-e_2|$; (this is consistent with 
an increase of level repulsion with increasing disorder in weak disorder limit).

Proceeding along the same lines, the above result can be generalized to more than two flat band. As reported in \cite{viddi} for case(e), an onset of disorder indeed gives rise  to three separated Gaussian level densities from 
three flat bands (see fig.(12) of \cite{viddi}).  

(ii)  {\it a flat band at the edge of a dispersive band}: For the combination of a flat band located at $e=0$ and a dispersive band with the level density $f_d(e)$, $R_1(e,Y_0)$ can be written as $R_1(e;Y_0)= {N\over 2} \;\left( \delta(e) + f_d(e,N) \right)$; the latter satisfies the normalization condition $\int_{-\infty}^{\infty} R_1(e) \; {\rm d}e =N$. The initial condition on $G(z,Y)$ now becomes
$G(z; Y_0) =  \lim_{\sigma \to 0} \; {N \over 2} \left( {\sqrt{2 \pi \sigma^2} \over \varepsilon \; e}  \; {\rm e}^{-{z^2 \over 2 \sigma^2}} + f_d(z,N) \right)$ where $f_d(e,N) = \lim_{\varepsilon \to 0} f_d(z,N)$. The latter along with 
eq.(\ref{g1}) then gives  

\begin{eqnarray}
G(z; Y) =  \lim_{\sigma \to 0} \; {N \over 2} \; \left({\sqrt{2 \pi \sigma^2} \over \varepsilon \; e}  \; {\rm e}^{-{(z- |Y-Y_0| \; G)^2 \over 2 \sigma^2}} + f_d((z- |Y-Y_0|\;G,N)\right)
\label{gd3}
\end{eqnarray}
%

For  $w <1$, the Gaussian term in the above equation can  be approximated as  ${\rm e}^{-{z^2 \over 2 \sigma^2}}$  (as  $f_d(e) \ll \delta(e)$, one can use the same approximation as in the single band) case. The calculation of 2nd term in eq.(\ref{gd3}) depends on the functional form of $f_d$. Writing 
$f_w(e,w,N) = \lim_{\varepsilon \to 0} f_d(z-(Y-Y_0)G,N)$, we have

\begin{eqnarray}
R_1(e; w) =   {N\over 2 \sqrt{2 \pi w^2}} \;{\rm e}^{-{e^2 \over 2 w^2} }  + {N\over 2} \;  f_w(e, w, N)
\label{gf4r}
\end{eqnarray}

 The effect of  disorder on the  level density for case (c) is displayed in figure 2 (also see figures 2(a) -5(a) of \cite{psf2}). Clearly for  $w <1$, $R_1/N$ is independent of disorder $w$ as well as size $N$ in the flat band but its behavior in dispersive band depends on the size. 


\subsection{Inverse Participation Ratio}

With spreading and merging of bands, the energy-dependence of ${\overline{\langle {\mathcal I}_2(e, \Lambda_I) \rangle}}$ plays an important role in the spectral statistics. The  initial condition needed to determine ${\overline{\langle {\mathcal I}_2(e, \Lambda_I) \rangle}}$  depends on the type of neighborhood. Here again we consider two examples:

(i) {\it Two flat bands}: The average IPR at $ Y > Y_0$ can  still be given by eq.(\ref{iq5}) with $J_0$ and $J_1$ given by eq.(\ref{j0}) and eq.(\ref{j2}). But now  the initial conditions on IPR and level density are  ${\mathcal I}_2(e,Y_0)= I_f \; \sum_{k=1}^2 \delta_{e e_k}$ and $R_1(e,Y_0)={N\over 2} \; \sum_{k=1}^2 \delta(e-e_k)$ respectively. To proceed further, one requires a prior knowledge of $R_1(e; Y)$ and ${\overline {\langle u  \rangle}}$(Y). For weak disorder $w <1$, $R_1$ is given by eq.(\ref{gf3r}). As the initial condition on local intensity in this case can be written as $\overline{\langle u\rangle}(e,Y_0) = u_0 \; \sum_{k=1}^2  \delta(e-e_k)$ with $u_0$ as a constant dependent on the eigenstates in two flat bands in the clean limit, eq.(67) of \cite{pslg} gives $\overline{\langle u\rangle}(e,y) = {u_0 \over \sqrt{2 \pi |Y-Y_0|}} \; \sum_{k=1}^2  \; {\rm exp} \left[-{(e-e_k)^2 \over {2 |Y-Y_0| }}\right]$  for a Gaussian Brownian ensemble.  
 The above on substitution in eq.(\ref{iq4}) gives
\begin{eqnarray}
 {\overline {\langle  {\mathcal I}_2  \rangle}}(e,\Lambda_I)  & \approx &  
 {1\over 2 \; R_1} \; \sqrt{{ 2 N  \over \Lambda_I  E_c^2}} \; 
\; \left[ N \; {\mathcal I}_f
   \; {\rm e}^{- \Lambda_I} \;  \sum_{k=1}^2 \; {\rm e}^{-{2 N\over \Lambda_I E_c^2}  \left(e-e_k-{ \Lambda_I E_c\over 4}\right)^2} + J_0 \right]
\nonumber \\
\label{iq8}
 \end{eqnarray}

where
\begin{eqnarray}
J_0  \approx { 3 \; u_0  \over w}  \; \sqrt{ \pi \Lambda_I \over N} \; 
 \sum_{k,l=1}^2 {\rm e}^{-{4 (e-e_k) \over E_c} } \;  {\rm e}^{-{1 \over 2 w^2}(e_l-e_k)^2 + {2\Lambda_I \over N}}  \; \Theta(e-e_k)
\label{j4}
\end{eqnarray}
with $\Lambda_I$ given by eq.(\ref{almi}).

For $e \approx e_1, e_2$, eq.(\ref{iq8}) again leads to (following the same reasoning as given below eq.(\ref{j3}) for case $e \sim 0$):

\begin{eqnarray}
{\overline {\langle  {\mathcal I}_2  \rangle}} (e)   
&\approx &  {6\; \pi\; u_0 \over N \; E_c} \;  
 {\rm e}^{{8 w^2 \over N E_c^2}}   \hspace{1in} \; {\rm for} \; e \sim e_1, e_2 
\label{jq8}
 \end{eqnarray}
Here the result for $e \sim e_2$ is obtained by neglecting the term ${\rm e}^{-{4 \over E_c} (e_2-e_1)}$; the approximation is valid for $E_c \propto N^{-\mu}$ and $e_2-e_1 >0$. Clearly the result for $e \sim e_1, e_2$ for weak disorder $w < {1\over 4} \; N^{1-2\mu}$ is same as eq.(\ref{aq9}) for a single band.

 The study \cite{nmg2} for case (d) with two flat bands at $e= e_2=-e_1=4$
gives ${\overline {\langle  {\mathcal I}_2  \rangle}} \propto N^{-0.83}$ with fractal dimension $D_2 \approx 2.49$ at $e=\pm 4$. Using $E_c \sim E_{th} \sim \Delta(e) \; N^{D_2/d}$ with $\Delta(e) \sim N^{-1}$, one has  $E_c \sim N^{-0.17}$, thus allowing the approximation ${\rm e}^{{8 w^2 \over N E_c^2}} \sim 1$ for $w < (1/4) N^{0.23}$. Clearly the average IPR near $e \sim \pm 4$  is independent from disorder but decreases with increasing size $N$: ${\overline{\langle{\mathcal I}_2  \rangle}} \sim N^{-0.83}$.  A same behavior was indicated by  \cite{nmg2} too.

With increasing disorder, the Gaussian level densities spread with their tails overlapping near  $e \sim {e_1+e_2\over 2}$ (middle of the gap region). An analysis of eq.(\ref{iq8}) in this region gives, for $e_2 > e_1$, 
 
\begin{eqnarray}
{\overline {\langle  {\mathcal I}_2  \rangle}}(e) 
&\approx &  {3\;\pi \; u_0 \over N \; E_c} \;  
 {\rm e}^{{8 w^2 \over N E_c^2}}  \; {\rm e}^{-2 (e_2-e_1) \over E_c} \;  {\rm e}^{(e_1-e_2)^2 \over 8 w^2}
\left[ 1 +  {\rm e}^{-(e_1-e_2)^2 \over 2 w^2} \right] 
\label{jq9}
 \end{eqnarray}

Note although the mean level spacing is $\Delta(e) \sim N^{-1}$ for both the regions,  $D_2$ is in general energy-dependent. As a result $E_c$ ($\sim \Delta \; N^{D_2/d}$) in the region $e \sim (e_1+e_2)/2$ is different from the centres (i.e $e \sim e_1, e_2$) of the Gaussian bands. This is also indicated by the numerical study in \cite{nmg2}) 
giving $E_c$ as $N^{-0.17} $ and $N^{-0.15}$  (with $D_2=2.49, 2.55$) for $e \sim 4$ and $e \sim 0$ respectively.   

As clear from eq.(\ref{jq9}), the average IPR has a different disorder dependence in the two energy ranges. At $e \sim (e_1+e_2)/2$, ${\overline {\langle  {\mathcal I}_2  \rangle}}(e,\Lambda_I)$ now decreases with increasing $w$ for $w\sqrt{2} < (e_2-e_1)$ but  increasing again for $w\sqrt{2} >(e_2-e_1)$. At 
$w \sqrt{2} \approx (e_2-e_1)$ and finite $N$, however the behavior at $e \sim (e_1+e_2)/2$ is almost analogous to that of $e \sim e_1, e_2$ if $E_c$ has a very weak N-dependence.  This is again consistent with numerical study in \cite{nmg2} for case (d) which indicates that IPR at  $w \approx 36$ and at $e \sim 4$ seems analogous to that of $e \sim 0$; note  $E_c$ values mentioned above indicate  a very slow variation in term ${\rm e}^{-2 (e_2-e_1) \over E_c}$ with $N$.

For $w >1$,  ${\overline {\langle  {\mathcal I}_2  \rangle}} $  can be obtained by substituting $\Lambda_I \approx {4 \;\ln |1-w^2| \over E_c^2}$ in eq.(\ref{iq8}). Proceeding  again as for $w <1$, one obtains 
\begin{eqnarray}
{\overline {\langle  {\mathcal I}_2  \rangle}} (e)   
&\approx &  {6\;\pi \; u_0 \over N \; E_c} \;  
 {\rm e}^{{8 \ln |1-w^2| \over N E_c^2}}  \qquad \; {\rm for} \; e \sim e_1, e_2 
\label{jjq8}    \\
&\approx &  {3\;\pi \; u_0 \over N \; E_c} \;  
 {\rm e}^{{16 \ln |1-w^2| \over N E_c^2}}  \; {\rm e}^{-2 (e_2-e_1) \over E_c} \;  {\rm e}^{(e_1-e_2)^2 \over 8 w^2}
\left[ 1 +  {\rm e}^{-(e_1-e_2)^2 \over 2 w^2} \right]   \quad \; {\rm for} \; e \sim (e_1+e_2)/2 
\nonumber \\
\label{jjq9}
\end{eqnarray}

Following similar steps, the above result can be generalized for cases with more then two bands. The study \cite{viddi} analyzes the disorder-sensitivity of ${\overline {\langle  {\mathcal I}_2  \rangle}}$  for case (e) (which has three flat bands for $\phi =\phi_0/2$ in the clean limit); their results again confirm the disorder-independence  in weak disorder limit (see figure 11 of \cite{viddi}). As the study \cite{viddi} does not analyze size-dependence of  ${\overline {\langle  {\mathcal I}_2  \rangle}}$, we are unable to compare our theoretical predictions with their results.

(ii) {\it Flat band at the edge of a dispersive band}: The initial conditions on $R_1$, IPR and $\langle u \rangle$ now become $R_1(e,Y_0)={N\over 2} (\delta(e) + f_d(e,N))$, 
${\mathcal I}_2(e,Y_0)= {\mathcal I}_f \; \delta_{e0} + {\mathcal I}_d \; \theta(e) $ and 
$\overline{\langle u\rangle}(e,Y_0) =  u_0 \; \delta(e) + u_1 \; \theta(e)$ 
 (with $\theta(e)=0, 1$ for $e <0$ and $e >0$ respectively, ${\mathcal I}_f, {\mathcal I}_d, u_0, u_1$ as constants dependent on the eigenstates properties in clean limit).
$R(e, Y)$ Is now given by eq.(\ref{gf4r})) and the above initial condition on $\overline{\langle u\rangle}$, eq.(69) of \cite{pslg}) leads to  $\overline{\langle u\rangle}(e,Y) \approx {u_0 \over \sqrt{2 |Y-Y_0| }} \; {\rm exp} \left[-{e^2 \over {2 \pi |Y-Y_0| }}\right] + u_d(e, w)$ with $u_d(e,w)$ as the local eigenfunction intensity in the dispersive band.   
Substitution of the above in eq.(\ref{iq5}) gives 
\begin{eqnarray}
 {\overline {\langle  {\mathcal I}_2  \rangle}}(e,\Lambda_I)  & \approx &  
  {1\over 2 R_1} \; \sqrt{{ 2 N  \over \Lambda_I  E_c^2}} \left[   N \;  {\rm e}^{-\Lambda_I }  \;  
\left( {\mathcal I}_f
\; {\rm e}^{-{2 N\over \Lambda_I E_c^2} \left(e-{\Lambda_I  E_c\over 4}\right)^2}  +  
 {\mathcal I}_d \; g_w(e) \right) + J_0 \right] .
\label{iq9}
\end{eqnarray}
with $g_w(e) = \int_0^{\infty} \; f_d(x) \;  {\rm e}^{-{2 N \over \Lambda_I E_c^2} \left(e -x-{\Lambda_I E_c \over 4} \right)^2} \; {\rm d}x$ and 
\begin{eqnarray}
J_0  \approx 3  \sqrt{\Lambda_I \over N w^2} \;\left[ u_0  \sqrt{\pi}  +B_1+B_2 + B_3 \right] \;  {\rm e}^{-{4 e \over E_c} + {2 \Lambda_I \over N}} \; . 
\label{j5}
\end{eqnarray}
where $B_1, B_2, B_3$ are integrals dependent on the level density and local intensity of the dispersive band:
\begin{eqnarray}
B_1 &=& {2 u_0 w \over E_c} \; \sqrt{\pi N\over \Lambda_I} \; \int_{-\infty}^{\infty} {\rm d}x \; f_w(x)  \; {\rm e}^{-{2 N x^2\over \Lambda_I E_c^2} +{4 x \over E_c}},\\
B_2 &=& N \int_{-\infty}^{\infty} {\rm d}x \; f_w(x) \; u_d(x) \; {\rm e}^{-{x^2\over 2 w^2} +{4 x \over E_c}},\\
B_3 &=& {\sqrt{2 \pi w^2} } \; \int_{-\infty}^{\infty} {\rm d}x \; f_w(x) \; u_d(x) \; {\rm e}^{4 x \over E_c},
\label{bb}
\end{eqnarray} 
For clarification, here we choose ${\mathcal I}_f=1, {\mathcal I}_d = {1 \over N}$, assuming all localized states in the flat band and all delocalized states in the dispersive band. 
Assuming $E_c \sim N^{-\mu}$ with $\mu >0$, here again $\Lambda_I \sim N^{2 \mu}$. 
Clearly, for large $N$ and finite $e$,  the contribution from the first  two terms in eq.(\ref{iq9}) 
 is  negligible which results in 
$ {\overline {\langle  {\mathcal I}_2  \rangle}}(e,\Lambda_I)  \approx 
  {1\over R_1} \; \sqrt{{ 2 N  \over \Lambda_I  E_c^2}} \; J_0$. Based on $e, w$, $J_0$ can further be simplified as follows

{\it Case $w < 1$, $ e \sim 0$: }
 $R(e, Y)$ for this case can again be approximated as $R_1(e) \approx {N \over \sqrt{2 \pi w^2}}$ (see eq.(\ref{gf4r})). Due to almost negligible contribution from the dispersive part near $e \sim 0$, $J_0$ in eq.(\ref{j5}) can again be reduced to the same form as in eq.(\ref{j3}), leading to ${\overline {\langle  I_2 \rangle}}(e,Y)$ independent of disorder but not of size:: 
\begin{eqnarray}
{\overline {\langle  {\mathcal I}_2  \rangle}}  \approx  {6 \; \pi \; u_0 \over N \; E_c}    \qquad e \sim 0, \; \; w < N^{1-2\mu}
\label{iq10}
\end{eqnarray}
As  mentioned below eq.(\ref{aq10}),  our numerical analysis of the two dimensional chequered board lattice gives $E_c \sim N^{-0.5}$ which implies ${\overline {\langle  {\mathcal I}_2  \rangle}} \sim N^{-1/2}$, an indicator of partially localized states.

{\it Case $w <1$, $ e > 0$: }  The dispersive contribution in $J_0$ in eq.(\ref{j5}) and $e$-dependence of $R_1$ in eq.(\ref{gf4r}) can be ignored. This results in, for $e > w \sqrt{2}$,  
\begin{eqnarray}
{\overline {\langle  {\mathcal I}_2  \rangle}}  \approx   {3\; \sqrt{2} \over N \; E_c \; w \; f_w(e)} \;  {\left[ u_0 \sqrt{\pi} + B_1+B_2 + B_3\right]} \; {\rm e}^{-{4 e \over E_c} } 
\label{iq11}
\end{eqnarray}

which implies an exponential decay away from the centre of the Gaussian band (for $E_c \propto N^{-\mu}$ for $\mu >0$).  This is again confirmed by our numerical analysis of the chequered board lattice (see figure 3). 

{\it Case $w >1$, $e \sim 0$:}
With $R_1(e \sim 0) \approx {N \over 2 \sqrt{2 \pi w^2}}+ {N \over 2} \; f_w(0,w)$, $\Lambda_I={4 \ln|1-w^2| \over E_c^2}$ and using eq.(\ref{j5}) for $J_0(e \sim 0)$, one has 
\begin{eqnarray}
{\overline {\langle  {\mathcal I}_2  \rangle}}  \approx   {6\;  \sqrt{\pi} \over N \; E_c} \;  {\left[ u_0 \sqrt{\pi} + B_1+B_2 + B_3\right] \over \left(1+w \sqrt{2 \pi}\; f_w(0,w)\right)} \;  {\rm e}^{ {8 \ln|1-w^2| \over N E_c^2}} 
\label{iq12}
\end{eqnarray}
The numerics for $d=2$ chequered board lattice (case (c)), with $w^2=10$ and near $e \sim 0$, gives $D_2 \approx 0.5$ implying $E_c \sim N^{-0.75}$ (see figure 4). The average IPR in this case is therefore approaching localized limit ${\overline {\langle  {\mathcal I}_2  \rangle}}  \sim N^{-0.25}$ and is also disorder dependent (see figure 3).

{\it  Case $w >1$, $e > 0$:} Here $\Lambda_I$ and $J_0$ can be given as in the previous case but now the Gaussian contribution to $R_1$ may not be ignored. This leads to

\begin{eqnarray}
{\overline {\langle  {\mathcal I}_2  \rangle}}  \approx   {6\; \sqrt{\pi} \over N \; E_c} \;  {\left[ u_0 \sqrt{\pi} + B_1+B_2 + B_3\right] \over \left({\rm e}^{-{e^2 \over 2 w^2}}+w \sqrt{2 \pi}\; f_w(e)\right)} \; {\rm e}^{-{4 e \over E_c}  + {8 \ln |1-w^2| \over N E_c^2}} 
\label{iq13}
\end{eqnarray}

\section{conclusion} 

Based on representation by a multi-parametric Gaussian ensemble, we derive a complexity parameter formulation of the ensemble averaged level density and inverse participation ratio
for disordered perturbed flat bands. Our results indicate  a disorder-insensitivity of these measures in weak disorder regime; this is consistent with  numerical results for a 2-dimensional chequered board lattice discussed in this paper and also with 3-dimensional  diamond lattice \cite{nmg2} as well as Aharonov-Bohm cages \cite{viddi}.  
A point worth emphasizing here is as follows: the results obtained here are applicable only for  those cases in which the diffusion of ensemble density can be represented by eq.(\ref{rhoy}), with initial state of diffusion corresponding to a macroscopic degeneracy  with localized eigen states.   A macroscopic degeneracy of energy levels (leading to peaked level density and localized eigenstates) can however arise in situations other than the flat bands \cite{su}. But that by itself does not ensure the applicability of our theoretical results.

The complexity parameter formulation of IPR helps in revealing an interesting tendency of the eigenfunction dynamics in the flat bands: the localization due to destructive interference of the highly degenerate flat band states seems to weaken with onset of disorder,  resulting in a partially localized wave-packet. It however becomes fully localized again beyond a critical disorder  due to impurity scattering. The variation of disorder thus leads to a  variation of the wave dynamics from localized $\to$ extended $\to$ localized phases; note however the wave-localization for weak and strong disorders has different origins. This in turn gives rise to  many questions e.g whether it is  possible to have a disorder driven transition in the flat bands? Is it different or analogous to disorder driven transitions in th dispersive bands and can it be defined in terms of a single scaling parameter? It is also relevant to know whether there exist a mobility edge in perturbed flat bands. We attempt to answer some of these questions in \cite{psf2}. As discussed in \cite{psf2}, the formulation not only helps us in search of criticality in perturbed flat bands, it also connects the latter to a wide range of other disordered systems.

In the present work, we have confined ourselves to the disorder perturbed bands. Previous studies have indicated many 
other system conditions which can play important role as perturbations e.g. symmetry or particle-interactions.   
Using a lattice with pentagon unit cell, the study \cite{gu2} indicates that a single particle dispersive band can be converted  into a flat band by an appropriate tuning of electron-electron interactions. As discussed in section III, 
 the complexity parameter formulation can also be applied to these case. But as the initial state in our $Y$ governed diffusion of the level density and IPR is chosen to be  a flat band (with dispersive band as the end of diffusion), the consistency of  our results with \cite{gu2} requires that a decrease of parameters $U_k$ leads to an increase of $Y$. To check this, we need an explicit formulation  of $Y$. Due to technical complications, the details of this case will be discussed elsewhere.

\vfill\eject

\oddsidemargin=-10pt
\begin{figure}
\centering
\includegraphics[width=1.\textwidth, height=1.2\textwidth]{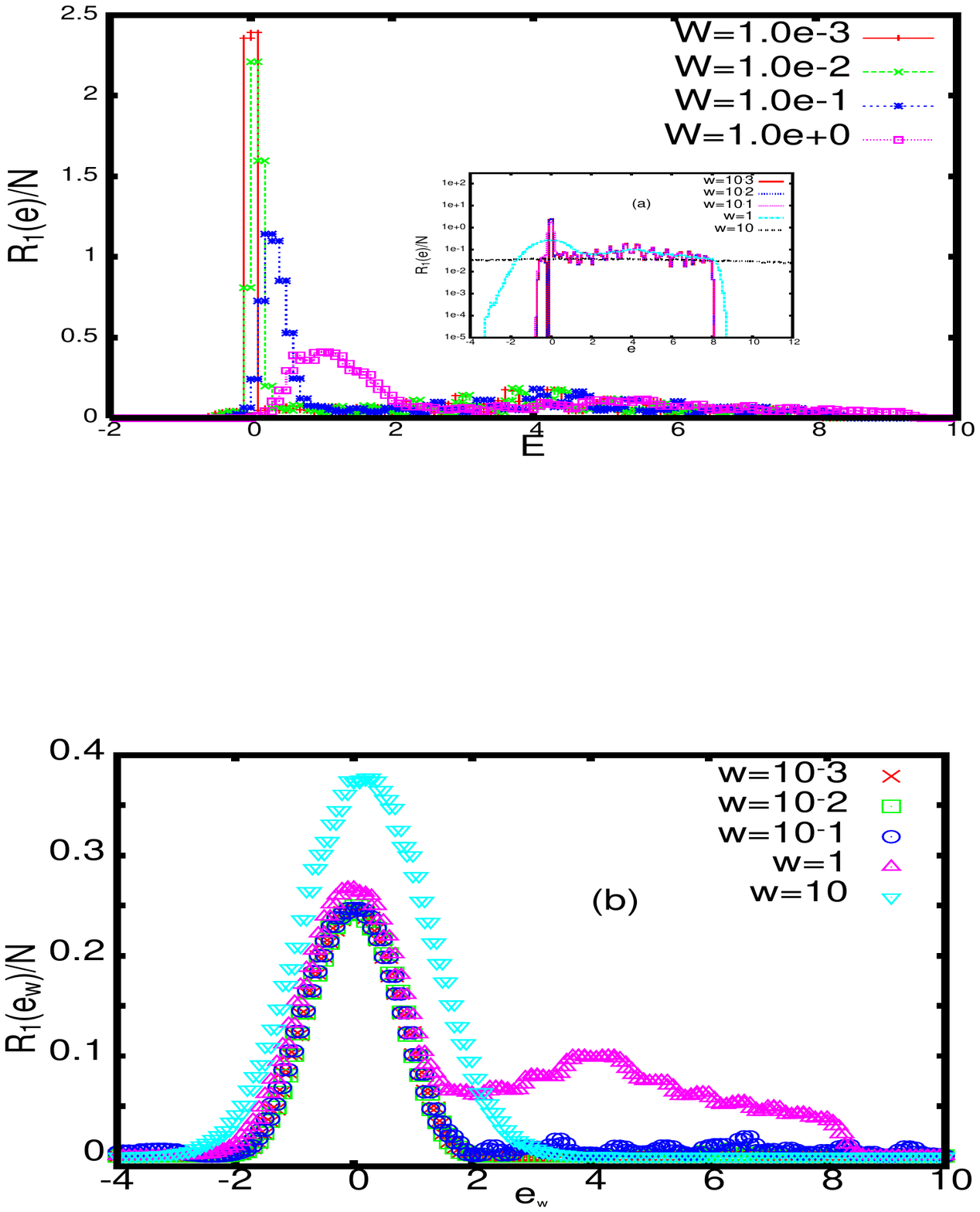} 
\vspace*{-60 mm}
\caption{ 
{\bf Ensemble averaged level density $N^{-1} \; R_1(e)$:} The figure displays the response of the level density of the  $2$-dimensional checkerboard lattice of linear size $L$ with on-site Gaussian disorder.  As mentioned in  example (c) of section II.A, the lattice consists of a flat band and a dispersive band; the choice of parameter $\epsilon=2t$ gives a flat band at $e=0$ and a dispersive band at $e >0$. (The numerical data is obtained for $t=1$, ${\mathcal M}=2$, $N=L^2$ with $L=34$): 
(a) $R_1(e)/N$ for both flat and dispersive bands subjected to various disorders $w$, with $W=w^2$ (the dispersive band behaviour is displayed more clearly in the inset on a semi-log scale), 
(b) $R_1(e)/N$ for the disorder perturbed flat band with respect to scaled energy $e_w=e/\sqrt{w}$ for various disorders. Clearly in terms of the rescaled energy, the level density in the perturbed flat band is disorder independent. For a clear visualization, the Gaussian fit for various disorder is not displayed here but is  shown in \cite{psf2} by the solid line fits in parts(a) of figures 2,3,4,5. The latter also confirms the size-independence of $R_1(e)/N$ for both all energy-ranges.  }
\label{fig2}
\end{figure}

\oddsidemargin=-10pt
\begin{figure}
\centering
\includegraphics[width=\textwidth, height=\textwidth]{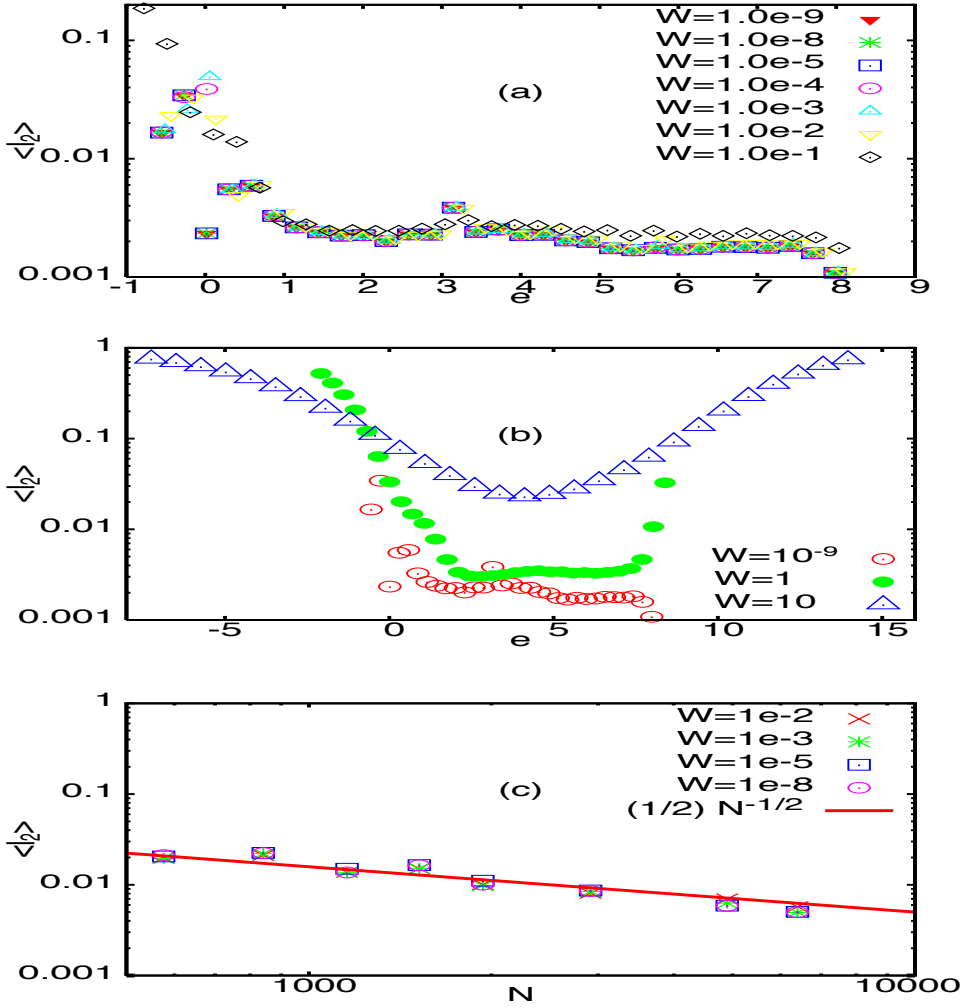} 
\vspace*{-20 mm}
\caption{
{\bf Ensemble averaged inverse participation ratio:  }
(a) energy-dependence of $\overline{\langle {\mathcal I}_2(e) \rangle}$  for a fixed size $L=34$, in weak disorder limit for energy ranges including both flat as well as dispersive band, (b) same as part (a) but now includes strong disorder cases too. 
Here ${\mathcal I}_2(e)$ is averaged over the ensemble as well as a small spectral window around each $e$ and $W=w^2$ with $w$ as the disorder.   
For weak disorders $w <1$, $\langle {\mathcal I}_2(e) \rangle$ for $e >0$ indicates a partially 
localized nature of wave-functions and is insensitive to disorder-strength $w$. 
However as shown in part (b), the disorder induced localization starts dominating the wave-function  for $w  \geq 1$. 
(c) size-dependence of $\langle I_2(e, N) \rangle$ for many $w$ for a fixed $e=0$ (middle of the flat band). Here only $10 \% $ of the 
eigenvectors from middle of the flat band are used in the analysis. 
The ensemble size is chosen so as to give approximately $10^3$
eigenfunctions for averaging for each $N$. 
The fit $ \langle I_2 \rangle \propto \sqrt{N}$ (equivalently $L^{-1}$ with $N=L^2$ for 
a 2-d chequered board lattice) suggest the 
multifractal exponent $D_2$ approximately 1; the $D_2$ numerics shown in figure 4 for different 
weak disorder strengths suggest $D_2 \sim 1.2$.   
}          
\label{fig3}
\end{figure}

\oddsidemargin=-10pt
\begin{figure}
\centering
\includegraphics[width=1.5\textwidth, height=1.5\textwidth]{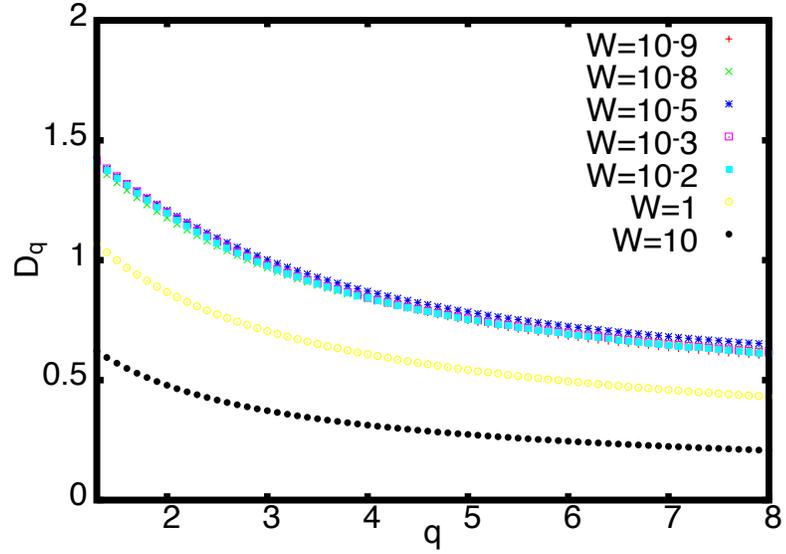} 
\vspace*{-6in}
\caption{ 
{\bf Disorder dependence of the fractal dimension $D_q$:}
The eigenvectors for the analysis are taken from the bulk of the flat band, with $W=w^2$. 
As shown later in figure 2-5 of \cite{psf2}, $D_q$ is size-independent too.  
}
\label{fig5}
\end{figure}

 \end{document}